\documentclass{aa}

\usepackage{graphicx}
\usepackage{txfonts}
\usepackage{CJKutf8}
\usepackage{longtable}
\usepackage{lscape}  
\usepackage{pdflscape} 
\usepackage{hyperref}
\usepackage{xcolor}
\usepackage{comment}

\begin{document}

   \title{CAMPOS II. The onset of protostellar disk substructures and planet formation}

   \author{{Cheng-Han Hsieh \begin{CJK}{UTF8}{bsmi}(謝承翰)\end{CJK}}
          \inst{1,2,3}
          \and
          H{\'e}ctor G. Arce\inst{2}
          \and 
          Mar{\'i}a Jos{\'e} Maureira\inst{4}
          \and 
          Jaime E. Pineda\inst{4}
          \and 
          Dominique Segura-Cox\inst{3}
          \and 
          Diego Mardones\inst{5}
          \and 
          Michael M. Dunham\inst{6}
          \and 
          Hui Li\begin{CJK}{UTF8}{gbsn} (李晖)\end{CJK}\inst{7}
          \and
          Stella S. R. Offner \inst{3,8}
          }

   \institute{The NASA Hubble Fellowship Program Sagan Fellow\\
              \email{chenghan.hsieh@utexas.edu}
         \and
             Department of Astronomy, Yale University, New Haven, CT 06511, USA
        \and 
        Department of Astronomy, The University of Texas at Austin, 2515 Speedway, Stop C1400, Austin, Texas 78712-1205, USA
        \and 
        Max Planck Institute for Extraterrestrial Physics, Gießenbachstraße 1, 85748, Garching bei München, Germany
        \and 
        Departamento de Astronomía, Universidad de Chile, Camino El Observatorio 1515, Las Condes, Chile
        \and Department of Physics, Middlebury College, Middlebury, VT 05753, USA        
        \and 
        Los Alamos National Laboratory, New Mexico 87545, USA
        \and National Science Foundation-Simons AI Institute for Cosmic Origins, The University of Texas at Austin, 2515 Speedway, Stop C1400, Austin, Texas 78712-1205, USA
             }

   \date{Submitted April 15, 2025 } 

\abstract{The 1.3\,mm CAMPOS survey has resolved 90 protostellar disks with $\sim$15 au resolution across the Ophiuchus, Corona Australis, and Chamaeleon star-forming regions. To address the fundamental question, `When does planet formation begin?', we combined the CAMPOS sample with literature observations of Class 0-II disks (bolometric temperature, $T_{\rm bol} \le 1900$ K), all mapped at 1.3 mm with resolutions ranging from 4 to 33 au. To investigate substructure detection rates as a function of bolometric temperature, we restricted the sample to disks observed at the 1.3 mm wavelength, with inclinations below 75$^\circ$, linear resolution $\le 20$\,au and resolved with at least four resolution elements ($\theta_{\rm disk}/\theta_{\rm res} \ge 4$). We also considered the effects of extinction correction and the inclusion of Herschel Space Telescope data on the bolometric temperature measurements to constrain the lower and upper limits of bolometric temperature for each source. We find that by $T_{\rm bol} \sim 200-400\,K$, substructure detection rates increased sharply to $\sim$60\,\%, corresponding to an approximate age of $0.2-0.4$\,Myr. No substructures are detected in Class 0 disks. The ratio of disk-averaged brightness temperature to predicted dust temperature shows a trend of increasing values toward the youngest Class 0 disks, suggesting higher optical depths in these early stages. Our statistical analysis confirms that substructures similar to those in Class II disks are already common by the Class I stage, and the emergence of these structures at $T_{\rm bol} \sim 200-400\,K$ could represent only an upper limit. Classifying disks with substructures into those with and without large central cavities, we find both populations coexisting across evolutionary stages, suggesting they are not necessarily evolutionarily linked. If protostellar disk substructures do follow an evolutionary sequence, then our results imply that disk substructures evolve very rapidly and thus can be present in all Class I/II stages and/or that they can be triggered at different times. }


\titlerunning{CAMPOS II. The onset of protostellar disk substructures}

   \maketitle

\section{Introduction}

One of the major challenges in the field of planet formation is understanding the relationship between circumstellar disk properties and the outcome of the planet formation process \citep{2023ASPC..534..501M}. The advent of high-resolution observations by the Atacama Large Millimeter/Submillimeter Array (ALMA) has fundamentally transformed our understanding of planet formation, unveiling intricate substructures within Class II disks (age $\gtrsim$1 Myr). These substructures manifest varied natures - some potentially sculpted by pre-existing planets, while others, such as dense rings, may act as nurseries for the formation of planetesimals and subsequent planet generations \citep{2015ApJ...808L...3A,2017ApJ...851L..23C,2019AJ....158...15P,2017ApJ...843..127D,2016ApJ...818...76J}. The prevalence of substructures in Class II protoplanetary disks has raised questions about how early these substructures form, and thus when the planet formation process begins \citep{2023ASPC..534..423B}.

Previous searches for disk substructures in Class 0/I disks have been limited and focused on the brightest and largest disks (e.g., \citealt{2019Natur.565..206S,2020Natur.586..228S,2020ApJ...902..141S,2021MNRAS.501.2934C,2023ApJ...951....8O,2023ApJ...951L...2L}). 
Our recent ALMA CAMPOS program has uniformly surveyed 18 Class 0, 31 Class I, 22 flat-spectrum, and 19 early Class II protostellar disks with a resolution of 15\,au in nearby star-forming regions within 200\,pc \citep{2024ApJ...973..138H}. Together with existing literature data, forming a total sample of 116 disks, we conducted a statistical study of the onset of disk substructures in embedded protostellar disks with a large sample and in different environments.

This paper is organized as follows: In Section ~\ref{sec:data} and ~\ref{sec:analysis}, we describe the substructure identification, mass calculation using the disk fluxes, and the construction of a homogeneous sample to search for the onset of disk substructures. 
We restricted the disk sample observed at the same 1.3\,mm wavelength, with inclination angle below 75$^\circ$ and resolved at least 4 beams. In Section ~\ref{sec:Results}, we present the main result of this paper: the substructure detection rate statistics as a function of evolutionary stage tracked by bolometric temperature. In addition, we discuss the existence of 2 populations of disk substructures. We discuss our results in Section ~\ref{sec:discussion}, and present our conclusion in Section ~\ref{sec:conclusion}.

\section{Data}
\label{sec:data}

\subsection{Observational data and the sample construction}
The detailed data reduction and the survey design for the CAMPOS survey are described in \citet{2024ApJ...973..138H}. We combined the CAMPOS uniform survey of 90 protostellar disks at $\sim 14-18$\,au in Corona Australis, Chamaeleon I \& II, Ophiuchus North, and Ophiuchus \citep{2024ApJ...973..138H} with protostellar disks from the high-resolution ($\sim7$\,au) eDisk survey \citep{2023ApJ...951....8O}, medium-resolution ($\sim28$\,au) ODISEA survey \citep{2019MNRAS.482..698C}, high-resolution ($\sim4$\,au) ODISEA survey \citep{2021MNRAS.501.2934C}, and high-resolution ($\sim4$\,au) DSHARP survey \citep{2018ApJ...869L..41A} to study the onset of disk substructures. We also included the high resolution ($\sim4$\,au) observation of HL Tau \citep{2015ApJ...808L...3A}. 

To construct a homogeneous sample to search for protostellar disk substructures, it is crucial to consider various observational biases. Detecting protostellar disk substructures requires spatially resolved disks. Previous studies of more evolved protoplanetary disks have shown that the detection rate of protoplanetary substructures strongly depends on the effective angular resolution ($\theta_D/\theta_{res}$), where $\theta_D$ is the diameter of the disk containing 90\% of the continuum emission and $\theta_{res}$ is the angular resolution of the observation. \citet{2023ASPC..534..423B} showed that if the effective angular resolution is below 3, substructures are detected in only 2\,\% of 256 protoplanetary disks. In contrast, substructures are detected in 52\,\% of the protoplanetary disks with intermediate effective angular resolution ($3 \le \theta_D/\theta_{res} \le 10$), and 95\,\% of the protoplanetary disks for ($\theta_D/\theta_{res} \ge 10$). To take into account this resolution bias and have a comparable criteria for the different sets of observations, we only analyze sources with $\theta_D/\theta_{res} \ge 4$\footnote{$\theta_{res}$ is estimated by taking the geometric mean of the FWHM of the major ($\theta_{\rm maj}$) and minor ($\theta_{\rm min}$) axes of the synthesized beam, that is $\theta_{res} = \sqrt{\theta_{\rm maj}\theta_{\rm min}}$.}.

Related to the resolution, we note that the absolute linear resolution naturally determines the smallest substructure size/scales that can be probed. The smallest detectable substructure scale is around half the beam size \citep{2012A&A...541A.135M}. We would use this to discuss the statistics of the size of substructures in the sample.

Besides the effective angular resolution and linear resolution, it is also important to compare observations at the same wavelength. The thermal and scattered continuum emission at a wavelength $\lambda$ is dominated by grains with a size of $\sim \lambda/2\pi$ \citep{1908AnP...330..377M}, meaning ALMA 1.3\,mm observations, such as the CAMPOS survey as well as DSHARP and eDISK, are more efficient at probing few 100 $\mu$m sized grains. In contrast, disk observations in the optical/near-infrared mainly probe much smaller sub-micrometer particles. To ensure a consistent survey probing the substructures from similar solids, we only include protostellar disks observed at 1.3\,mm wavelength by ALMA. Finally, the occurrence rate of substructures is also influenced by the disk's inclination angle. Due to the increasing optical depth and reduced emitting size, searching for disk substructures in edge-on disks is extremely challenging. While most of the high-resolution observations towards the more evolved Class II disks avoided highly inclined sources (e.g., DSHARP), the population of Class 0/I disks in nearby regions is more limited, and thus samples of very high-resolution observations (e.g., eDisk) contain a significant amount of highly inclined disks. To take this into account, we exclude all disks in the sample with inclination angles larger than 75$^\circ$.\footnote{We assume thin circular disks and use the major-to-minor axis ratio of the dust continuum data to compute the inclination angle ($i$), that is $i = sin^{-1}(R_{\rm min}/R_{\rm maj})\times 180^\circ/\pi$.}

\subsection{Substructure criteria}

In this paper, we labeled disks with substructures if the disk has rings, gaps, a central cavity, or spirals. Sources with asymmetric intensity profiles but lacking annular or spiral structures (e.g., IRAS 16293-2422B, \citealt[]{2021MNRAS.508.2583Z,2024A&A...682A..56Z,2024ApJ...973..138H}) are counted as disks without substructures.

We note that following the above criteria, all of the disks considered as with substructures in this work have already been confirmed as such in separate published papers (CAMPOS: \citet{2024ApJ...973..138H}, eDisk: \citet{2023ApJ...951....8O}, ODISEA: \citet{2019MNRAS.482..698C,2021MNRAS.501.2934C}, DSHARP: \citet{2018ApJ...869L..41A}.), and all correspond to substructures that are immediately visible in the 1.3 mm image. Sources that show small variations as a function of distance in $uv-$space or small variation in the intensity radial profile but appear smooth in the image plane are counted here as disks without substructures (e.g., \citealt[]{2023AJ....166..184M}). This is a conservative approach, and we may miss some sources with borderline detections, introducing a bias against younger disks with smaller and lower-contrast (shallower) substructures (e.g., \citealt[]{2024arXiv240720074M}). The implications of this will be further discussed in Section \ref{sec:discussion}.

\section{Analysis}
\label{sec:analysis}

\subsection{Disk fluxes and sizes}

In CAMPOS I, we presented a study on the evolution of protostellar disk radii \citep{2024ApJ...973..138H}. Both disk sizes and fluxes were derived from a 2D Gaussian fit with the CASA \textit{imfit} task \citep{2007ASPC..376..127M}. Disk sizes are calculated as the 2$\sigma$ size of the deconvolved major axis, which corresponds to approximately the radius containing 90\% of the flux R$_{90}$. For sources that exhibit emission profiles that deviate significantly from a Gaussian shape, due to disk substructures or have poor CASA \textit{imfit} results due to low S/N, we used the $5\sigma$ contour in the Briggs 0.5 weighted maps to measure the radius and flux \citep{2024ApJ...973..138H}\footnote{ These sources are DoAr 20, Oph-emb-20, CFHTWIR-Oph 79, IRS 2, SMM 2, ISO-ChaI 101, ISO-ChaI 207, shown in Table \ref{table:substructure_data}.}. We present the CAMPOS flux measurements in Table \ref{table:substructure_data}. Note that if a CAMPOS source was also observed at the same frequency (ALMA Band 6) but with higher angular resolution by other surveys, we report the protostellar disk flux and size from the literature instead. This ensures that for each disk presented in Table \ref{table:substructure_data}, the disk flux and radius are measured using the highest angular resolution ALMA data available.

\subsection{Deriving the mass of protostellar disks }

We adopted the method outlined in the ODISEA survey of Ophiuchus protoplanetary disks to convert the 1.3\,mm disk flux into dust mass. This method assumes optically thin emission, a constant temperature of 20\,K and an opacity coefficient of $\kappa_\nu= (\nu/100 \, {\rm GHz})$\,cm$^2$\,g$^{-1}$ \citep{2019ApJ...875L...9W}. We then assumed a gas-to-dust ratio of 100 to obtain the disk mass \citep{2023ASPC..534..423B}. We caution that disks might not be fully optically thin, particularly at the early stages \citep[e.g.,][]{2017ApJ...840...72L,2021MNRAS.508.2583Z,2021MNRAS.501.1316L,2022ApJ...941L..23M}, but also at the Class II stage \citep{2018ApJ...865..157A,2021MNRAS.506.2804T,2024ApJS..273...29C}. In this case, the estimated masses can be conservatively considered a lower limit. The protostellar disk mass measurement is for better comparison with other works and as a guide, and it does not affect the analysis of the onset of disk substructures.

\subsection{Bolometric temperature as a disk evolutionary tracer}
\label{sec:dis_on_set_caveat}

We cross-matched all protostellar disks with the Young Stellar Objects catalog from the Spitzer Space Telescope ``cores to disks" (c2d) and ``Gould Belt" (GB) Legacy surveys \citep{2015ApJS..220...11D}, as well as the Extension of HOPS Out to 500 ParSecs (eHOPS) catalog (Riwaj Pokhrel, private communication), to obtain the bolometric temperature ($T_{\rm bol}$), which serves as a proxy for relative evolutionary age for embedded protostellar systems (see \autoref{Appendix_B}). The eHOPS catalog contains 1-850 $\mu$m Spectral Energy Distributions (SEDs) assembled from Two Micron All Sky Survey (2MASS), Spitzer, Herschel Space Telescope (Herschel), Wide-field Infrared Survey Explorer (WISE), and James Clerk Maxwell Telescope (JCMT)/SCUBA-2 data. The first eHOPS paper for the Serpens and Aquila molecular clouds was published by \citet{2023ApJS..266...32P}. For all other clouds, the SED and protostellar system properties are available in the \href{https://irsa.ipac.caltech.edu/data/Herschel/eHOPS/overview.html}{NASA/IPAC Infrared Science Archive}\footnote{ https://irsa.ipac.caltech.edu/data/Herschel/eHOPS/overview.html}. For sources without a counterpart in both eHOPS and \citet{2015ApJS..220...11D}, we adopted the $T_{\rm bol}$ from the literature. Given that we want to investigate the emergence of substructures, we only included samples with $T_{\rm bol} \le 1900$ K. The final list of sources with $T_{\rm bol} \le 1900$ K and an inclination angle less than $75^\circ$ is shown in Table \ref{table:substructure_data}. In Table \ref{table:CAMPOS_Tbol}, we present the alternative names for each source in the CAMPOS survey for the  
Corona Australis, Chamaeleon I \& II, Ophiuchus North, and Ophiuchus sources. The cross-matched results for the ODISEA survey are presented in Table \ref{table:ODISEA2019_Tbol}.

In this paper, we consider the bolometric temperature measurements from both the eHOPS catalog and the \citet{2015ApJS..220...11D} catalog to trace the onset of protostellar disk evolution. Compared to \citet{2015ApJS..220...11D}, the eHOPS catalog provides more complete spectral energy distributions (SEDs) from $1$ to $850\,\mu$m. However, the infrared portion of the SED is not corrected for extinction \citep{2023ApJS..266...32P}. To maintain consistency with the eHOPS catalog, we use the bolometric temperatures without extinction correction from \citet{2015ApJS..220...11D} for sources without a counterpart in eHOPS. The extinction correction applied to the infrared data increases the infrared flux and thus further increases the bolometric temperature \citep{2015ApJS..220...11D}. Results based on this correction represent the upper age limit for the onset of disk substructures as longer-wavelength Herschel data are not included from the SED fitting to compute $T_{\rm bol}$. Therefore, results based on the eHOPS bolometric temperatures represent a lower age limit for the onset of disk substructures. On the other hand, to investigate the upper age limit for the onset of disk substructures, we adopt the extinction-corrected bolometric temperature from \citet{2015ApJS..220...11D}.

The bolometric temperature measured from the eHOPS catalog and \citet{2015ApJS..220...11D} differs from source to source. The decrease in $T_{\rm bol}$ is not uniform when including Hershel data into the \citet{2015ApJS..220...11D}. It depends both on the shape of the SED and on how well the SED was sampled by \citet{2015ApJS..220...11D}. While Herschel photometry was not available for the SED fits in \citet{2015ApJS..220...11D}, many of the protostars had 450\,$\mu$m observations from SCUBA, some had 350\,$\mu$m observations from SHARC-II on Caltech Submillimeter Observatory, and some brighter objects had Infrared Astronomical Satellite (IRAS) or Infrared Space Observatory (ISO) far-infrared photometry. Spitzer 70\,$\mu$m photometry was also available for nearly the full Spitzer-GB sample, but at a factor of 3 lower spatial resolution than Herschel, resulting in confusion for protostars with close neighbors. As a result, some $T_{\rm bol}$ values change by a large amount ($\sim$40\,\%), while some change by a very small amount ($\lesssim$10\,\%). We present the $T_{\rm bol}$ from both eHOPS and \citet{2015ApJS..220...11D} to quantify the range for the onset of protostellar disk substructures in Table \ref{table:substructure_data}.

\section{Results}
\label{sec:Results}

\subsection{Protostellar disk substructures versus $T_{\rm bol}$ and disk properties}
\label{sec:on_set_disk_substructure}

\begin{figure*}[tbh!]
    \includegraphics[width=.99\textwidth]{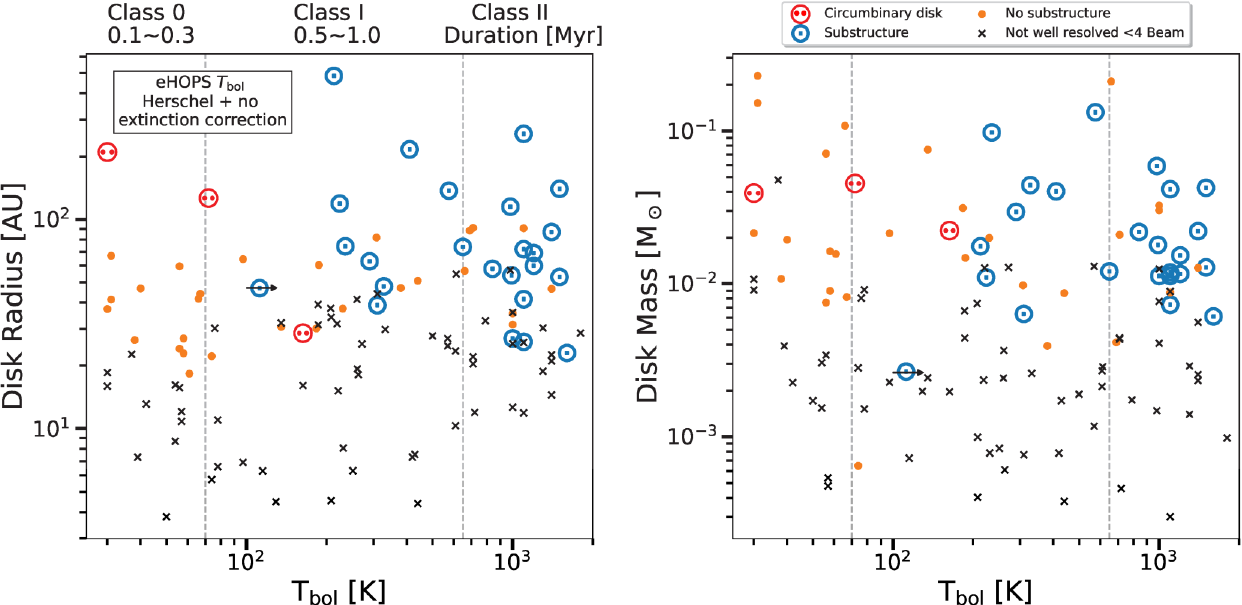}
    \caption{Protostellar disk substructure detections as a function of disk radius, mass, and bolometric temperature. The latter is taken from the eHOPS catalog. The gray dashed lines indicate the approximate boundary between the Class 0 and Class I, and between the Class I and Class II evolutionary phases \citep{1995ApJ...445..377C}, along with the corresponding expected duration of each Class \citep{2015ApJS..220...11D}. While the $T_{\rm bol}$ is not linear with age, in general the Class 0 sources have $T_{\rm bol} <70$\,K, Class I sources have $70 < T_{\rm bol} < 650$\,K, and Class II and Class III sources have $T_{\rm bol}>650$\,K \citep{2021A&A...653A.117B}. 
    The black arrow is used to indicate that the particular source has also been classified as Class II in the literature \citep{2023ApJ...958...20N}.}  
\label{fig:Substructure_evolution}
\end{figure*} 

\begin{figure*}[tbh!]
    \includegraphics[width=1\textwidth]{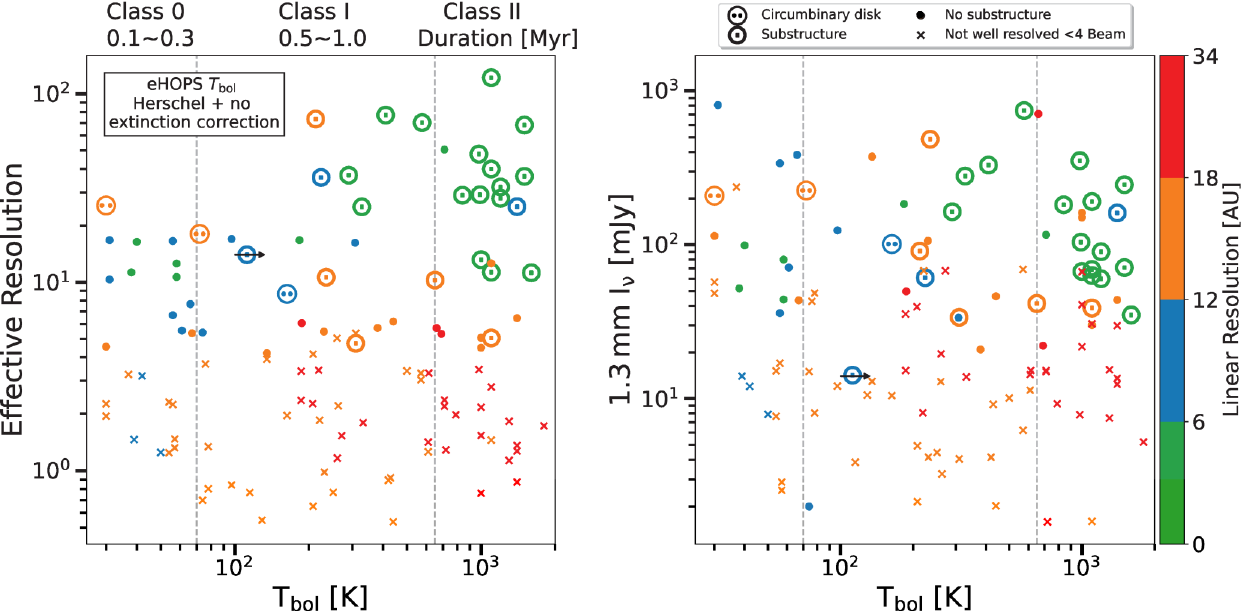}
    \caption{Substructure detection as a function of effective resolution, 1.3 mm flux and bolometric temperature, the latter from the eHOPS catalog. The black arrow is used to indicate that the particular source has also been classified as Class II in the literature \citep{2023ApJ...958...20N}.} 
\label{fig:Flux_Tbol}
\end{figure*}

\begin{figure*}[tbh!]
    \includegraphics[width=1\textwidth]{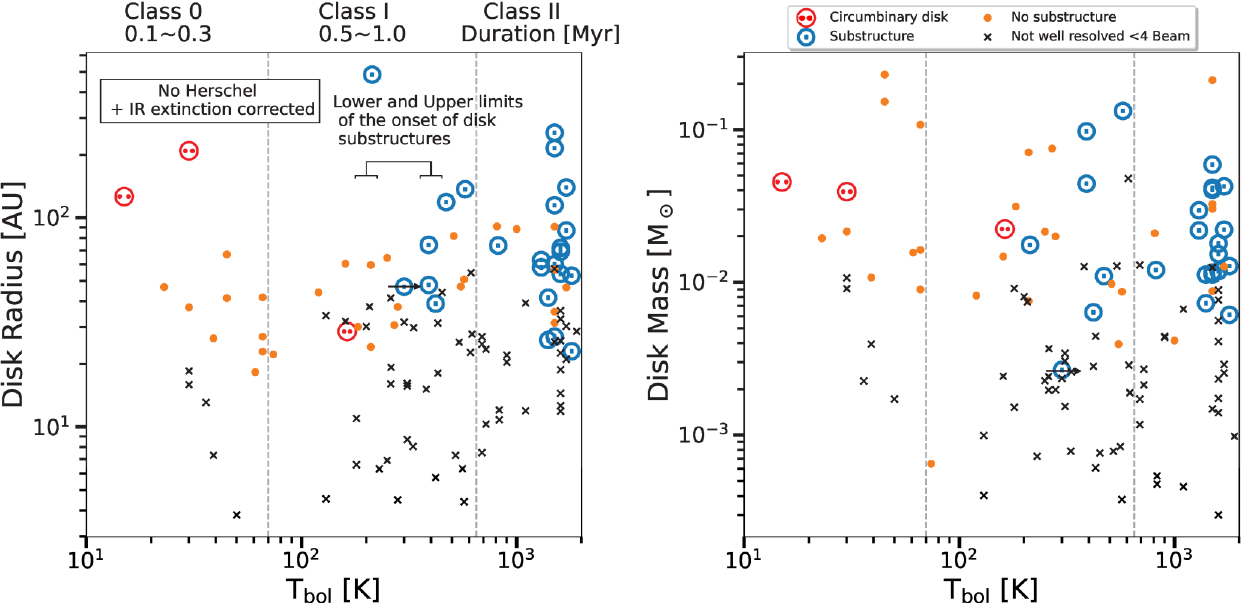}
    \caption{The upper limit of the onset of disk substructures. Same as Figure \ref{fig:Substructure_evolution}, but the bolometric temperature is derived from the SED fitting without the Herschel Space Observatory data ($70-500\,\mu$m) \citep{2015ApJS..220...11D}. Extinction correction is applied to the infrared data, which further increases the bolometric temperature \citep{2015ApJS..220...11D}. 
    The effective resolution and the ALMA Band 6 1.3\,mm flux of protostellar disks are color-coded and shown in different symbols according to the presence or absence of substructure. Compact disks with diameters smaller than 4 beams are indicated by crosses. The black arrow is used to indicate that the particular source has also been classified as Class II in the literature \citep{2023ApJ...958...20N}.
    } 
    \label{fig:Substructure_evolution_upper_limit}
\end{figure*}

\begin{figure*}[tbh!]
    \includegraphics[width=.99\textwidth]{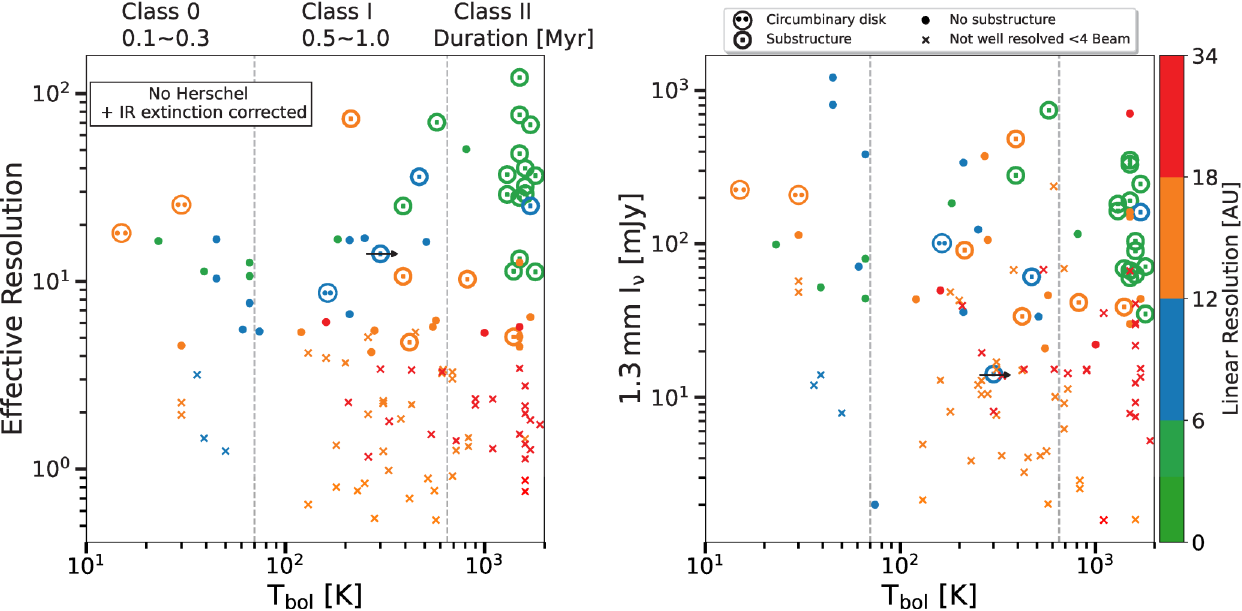}
    \caption{The upper limit of the onset of disk substructures. Same as Figure \ref{fig:Flux_Tbol}, but the bolometric temperature is derived from the SED fitting without the Herschel Space Observatory data ($70-500\,\mu$m) \citep{2015ApJS..220...11D}. Extinction correction is applied to the infrared data, which further increases the bolometric temperature \citep{2015ApJS..220...11D}. The crosses represent compact disks that are not well-resolved with diameters less than 4 times the angular resolution of the observation (beams). The lower limit of the onset of the disk substructures at 200\,K found in Figure \ref{fig:Substructure_evolution} is also labeled. The black arrow is used to indicate that the particular source has also been classified as Class II in the literature \citep{2023ApJ...958...20N}.
    }  
\label{fig:Flux_Tbol_upper_limits}
\end{figure*} 

Figure \ref{fig:Substructure_evolution} illustrates the radius and mass of the protostellar disks as a function of bolometric temperature ($T_{\rm bol}$) derived from the eHOPS catalog. The black dashed lines indicate the $T_{\rm bol}$ that marks the approximate boundary between the Class 0 and Class I, and between the Class I and Class II evolutionary phases \citep{1995ApJ...445..377C}, along with the corresponding expected duration of each Class \citep{2015ApJS..220...11D}. For reference, we also include the sources that do not comply with the effective resolution criteria ($\theta_D/\theta_{\rm res} \ge 4$). The latter sources have, in general, smaller sizes and fluxes/masses. 

The earliest substructure observed, IRAS 04169+2702, corresponds to a disk with an approximate age of $0.1-0.2$ Myr ($T_{\rm bol}\sim 163$\,K). If we consider the lower limit of $T_{\rm bol}$ of CFHTWIR-Oph 79 ($T_{\rm bol}\sim 110$\,K), then the earliest substructure could be even younger.\footnote{The lower limit of the $T_{\rm bol}$ from this source corresponds to the calculation from eHOPS. On the other hand, in the literature, this source has been classified as a Class II, hence the lower limit.} 

We find no detection of substructure in any of the Class 0 disks in the sample. Figure \ref{fig:Flux_Tbol} shows the effective resolution and flux of the disks with and without substructures as a function of bolometric temperature, color-coded by the linear resolution. We can observe that the lack of substructure in Class 0 disks is evident across resolutions from 4 to 18 au.

Figure \ref{fig:Substructure_evolution_upper_limit} and \ref{fig:Flux_Tbol_upper_limits} show similar plots as Figure \ref{fig:Substructure_evolution} and \ref{fig:Flux_Tbol} but considering the upper limit to bolometric temperatures from \citet{2015ApJS..220...11D}. In that case, the earliest substructures discussed above correspond to $T_{\rm bol}$ of 163 and 300\,K, for IRAS 04169+2702\footnote{For IRAS 04169+2702, the $T_{\rm bol}$ is adopted from \citet{2023ApJ...951....8O}.} and CFHTWIR-Oph79, respectively.

\subsection{Disk substructure detection rates}

\begin{figure*}
    \includegraphics[width=12cm]{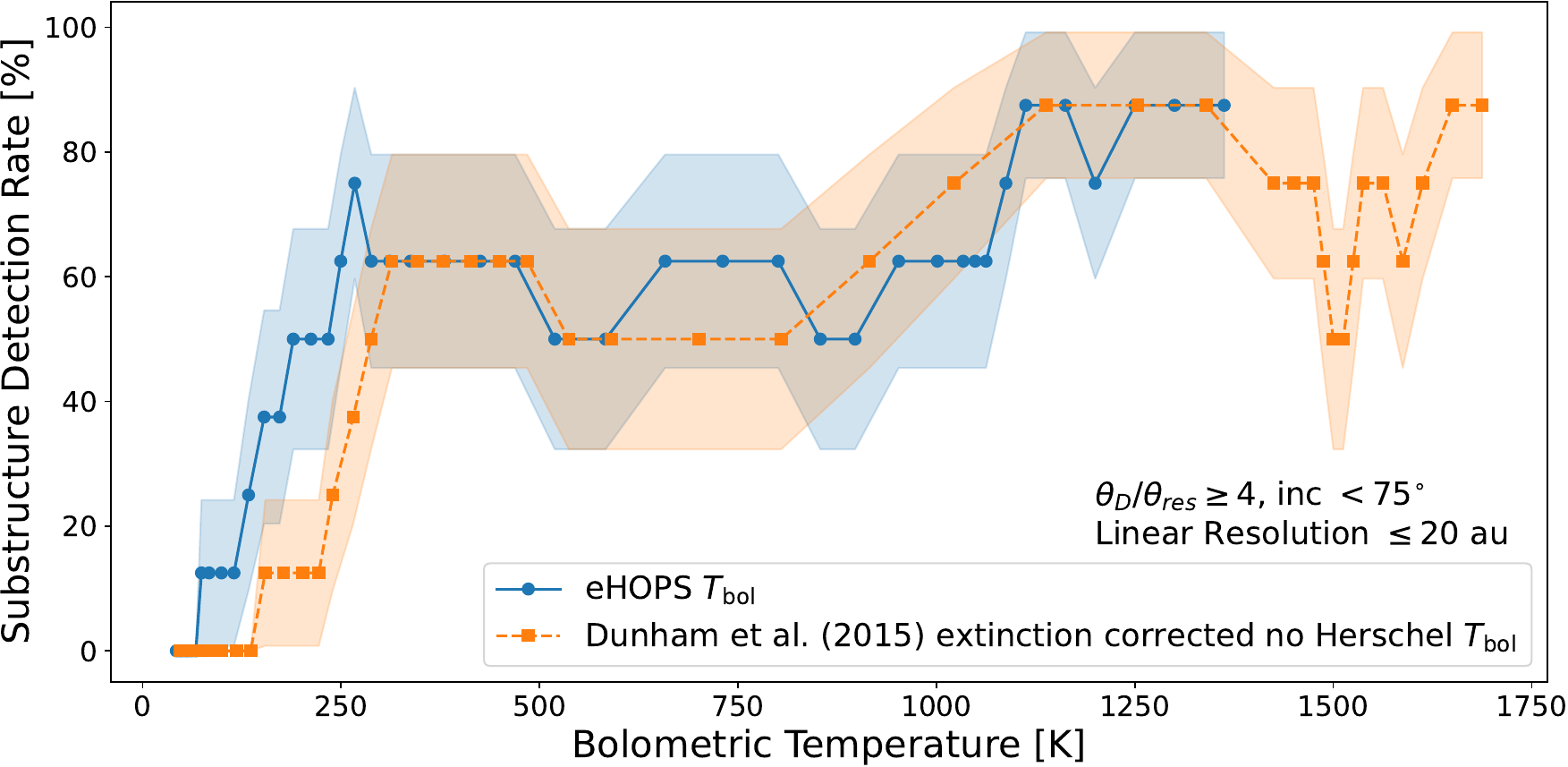}
    \caption{Disk substructure detection rates versus the evolutionary stage traced by bolometric temperature. The blue and orange represent the lower and upper limits with the bolometric temperature ($T_{\rm bol}$) measured by the eHOPS (with Herschel, but not extinction corrected) and by \citet{2015ApJS..220...11D} (no Herschel, but extinction corrected), respectively. The bin size of 8 disks is used to compute the substructure detection rates. The shaded region represents the $1\sigma$ uncertainty from counting statistics. All 51 protostellar disks are observed by ALMA Band 6 at 1.3\,mm wavelength, have a disk diameter greater than 4 times the resolution ($\theta_D/\theta_{res} \ge 4$), with an inclination angle less than 75\,$^\circ$, and linear resolution less than 20\,au.}  
\label{fig:Substructure_detection_rate}
\end{figure*}

We present the protostellar disk substructure detection rates as a function of bolometric temperature in Figure \ref{fig:Substructure_detection_rate}. Each disk is assigned a value of 1 if it exhibits substructures and 0 if it does not. We compute a moving average with a window size of 8, where the kernel slides through $T_{\rm bol}$ and calculates the mean for every group of 8 disks. This means represents the substructure detection rate. Due to the fixed window size, the detection rates are sampled at discrete levels of 0/8, 1/8, 2/8, ..., up to 8/8 — effectively producing 8 bins in the substructure detection rate space.

All 51 protostellar disks in Figure \ref{fig:Substructure_detection_rate} are observed by ALMA Band 6 at 1.3\,mm wavelength, have a disk diameter greater than 4 times the resolution ($\theta_D/\theta_{res} \ge 4$), with an inclination angle less than 75\,$^\circ$, and linear resolution less than 20\,au. We adopt the moving average with a window size of 8 to ensure enough sampling resolution across $T_{\rm bol}$ and substructure detection rate space. The blue and orange represent the case when the bolometric temperature is adopted from eHOPS catalog and the $T_{\rm bol}$ from \citet{2015ApJS..220...11D}, respectively. The shaded region represents the uncertainty from the binomial counting uncertainty ($\sigma_b$), which is calculated as
\begin{equation}
\sigma_b=\sqrt{\frac{r(1-r)}{n}},
\end{equation}
where $r$ is the substructure detection rate, and $n=8$ is the total number of sources in the window. The substructure detection rate adopting the \citet{2015ApJS..220...11D} $T_{\rm bol}$ in general agrees with the results adopting the eHOPS $T_{\rm bol}$, with a delay of $\sim 200$\,K in the rapid rise of disk substructures detection rate at $T_{\rm bol}\sim 200-400$\,K. 

Figure \ref{fig:Substructure_detection_rate} shows that the rate quickly reaches $\sim$60\% at $T_{\rm bol}$ of 200--400\,K, clearly indicating that substructure during the Class I stage is not rare. The Class I protostellar disk substructure detection rate in Figure \ref{fig:Substructure_detection_rate} is also consistent with the $\sim$50\% substructure detection rate in a sample of 62 Class II protoplanetary disks observed with an effective resolution $3 < \theta_D/\theta_{res} \le 10$ \citep{2023ASPC..534..423B}, strongly suggesting that the current detection rate of 60\% from $T_{\rm bol}$ 200--400\,K, is only a lower limit given the more limited resolution of the younger disk sample (see Figure \ref{fig:Flux_Tbol}). The high Class I disk substructure detection rate supports that the onset of disk substructures does not occur when disks transit into the Class II phases, but instead firmly into the Class I stage and possibly earlier (Section \ref{sec:discussion}).

\begin{figure*}[tbh!]
    \includegraphics[width=1\textwidth]{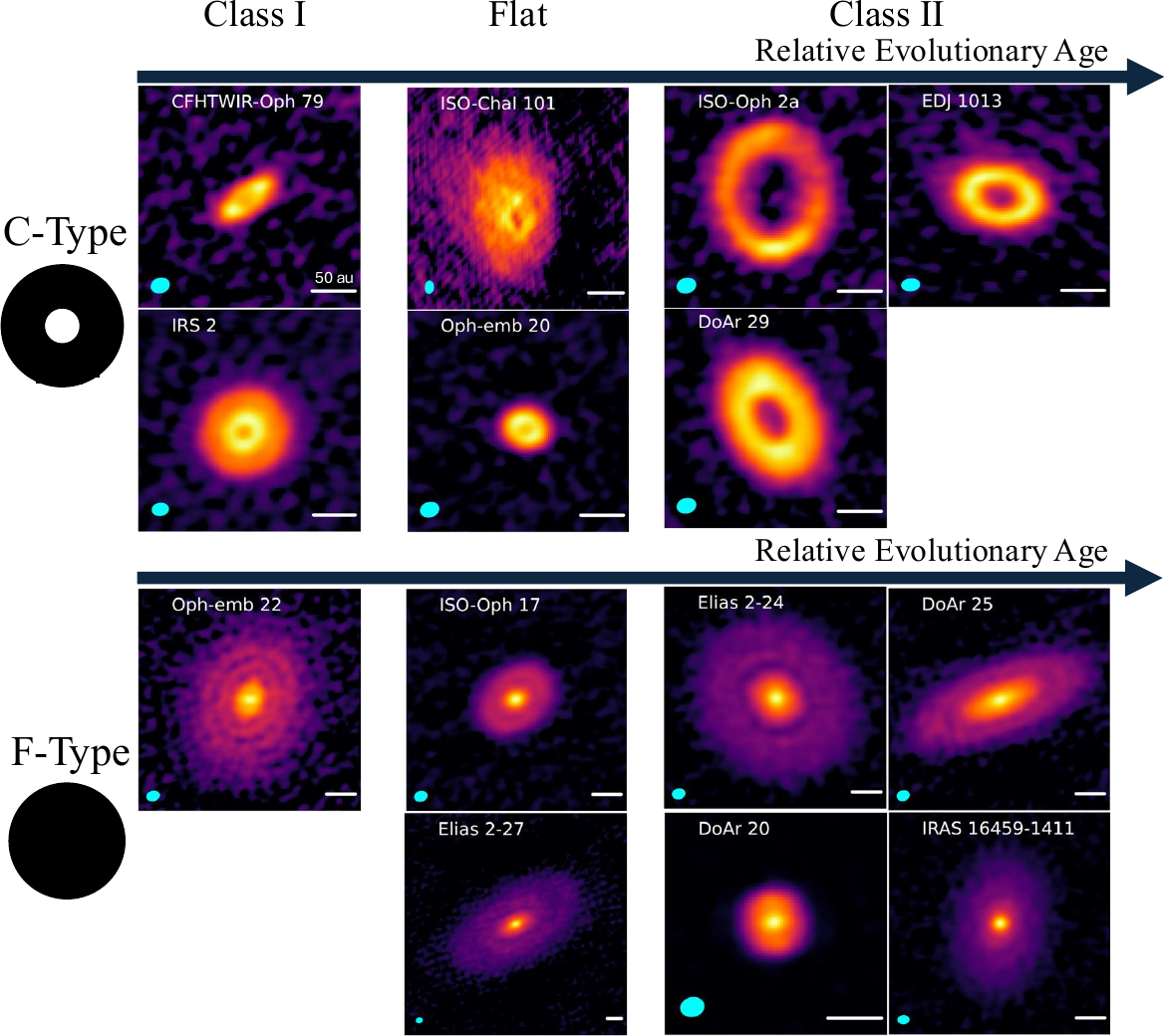}
    \caption{The continuum images of ``C-type" and ``F-type" disks as a sequence of evolutionary class. Both types of disk substructures can be detected throughout the Class I to Class II phases. } 
\label{fig:2_population}
\end{figure*}

\begin{figure*}[tbh!]
    \includegraphics[width=1\textwidth]{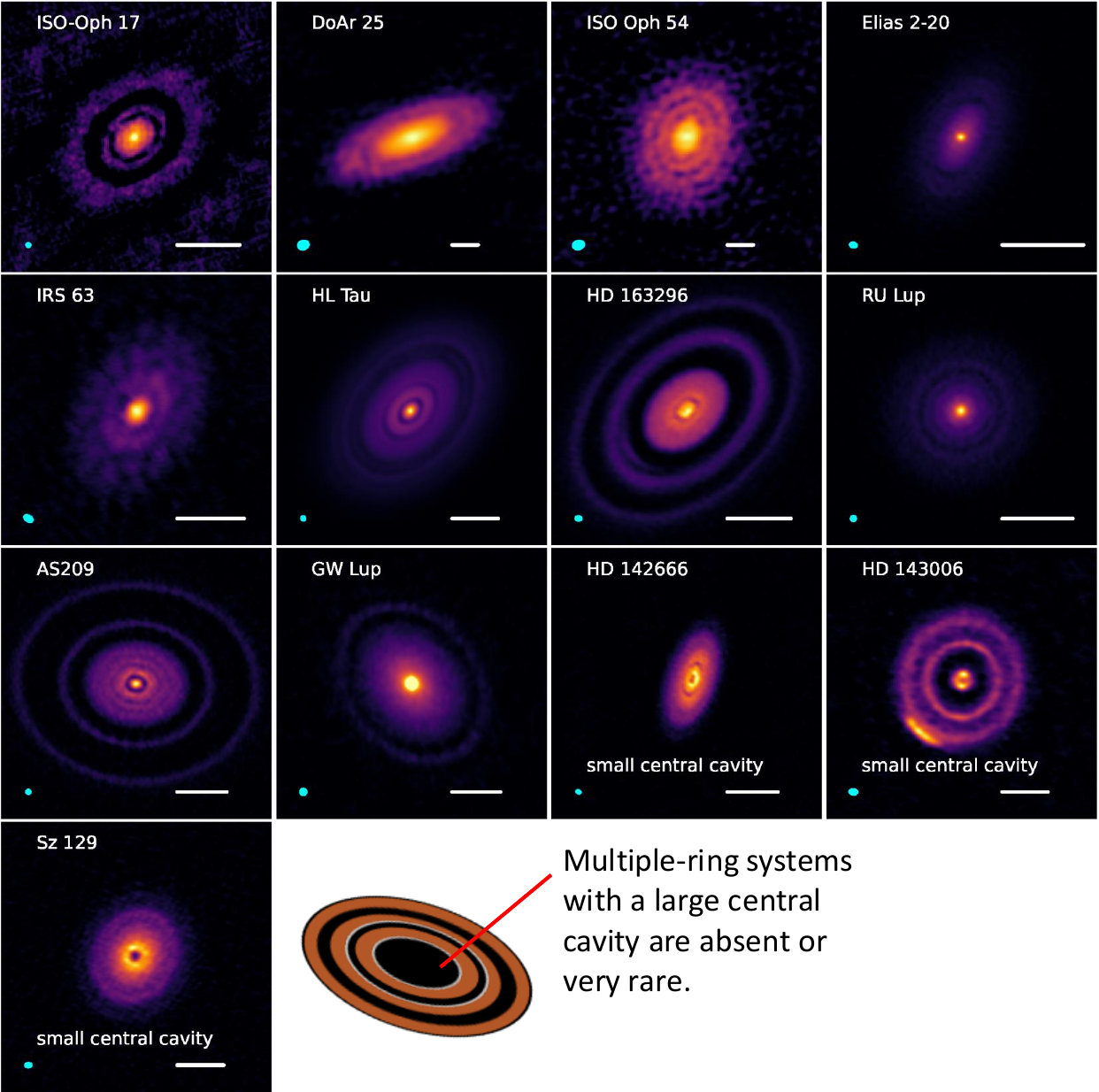}
    \caption{Dust continuum images of disks with multiple rings in our sample and the literature. Most of the multiple-ring systems are centrally filled (F-type) \citep{2018ApJ...869L..41A,2021MNRAS.501.2934C,2023ApJ...951....8O,2024ApJ...973..138H,2019MNRAS.482..698C,2015ApJ...808L...3A,2023Natur.623..705S}. The lack of a large central cavity ($\ge$ 25\,\% of the disk diameter) implies that a centrally filled disk is a general feature of multiple-ring systems. The cyan circle represents the beam size. The white lines mark the scale of 50\,au. Note that HD 163296, RU Lup, AS 209, GW Lup, HD 142666, HD 143006, and Sz 129 either do not have reliable bolometric temperature ($T_{\rm bol}$) measurements or their $T_{\rm bol} \ge 1900$\,K, thus are not included in this study.} 
\label{fig:Multiple_rings}
\end{figure*}

\begin{figure*}[tbh!]
\centering
    \includegraphics[width=0.88\textwidth]{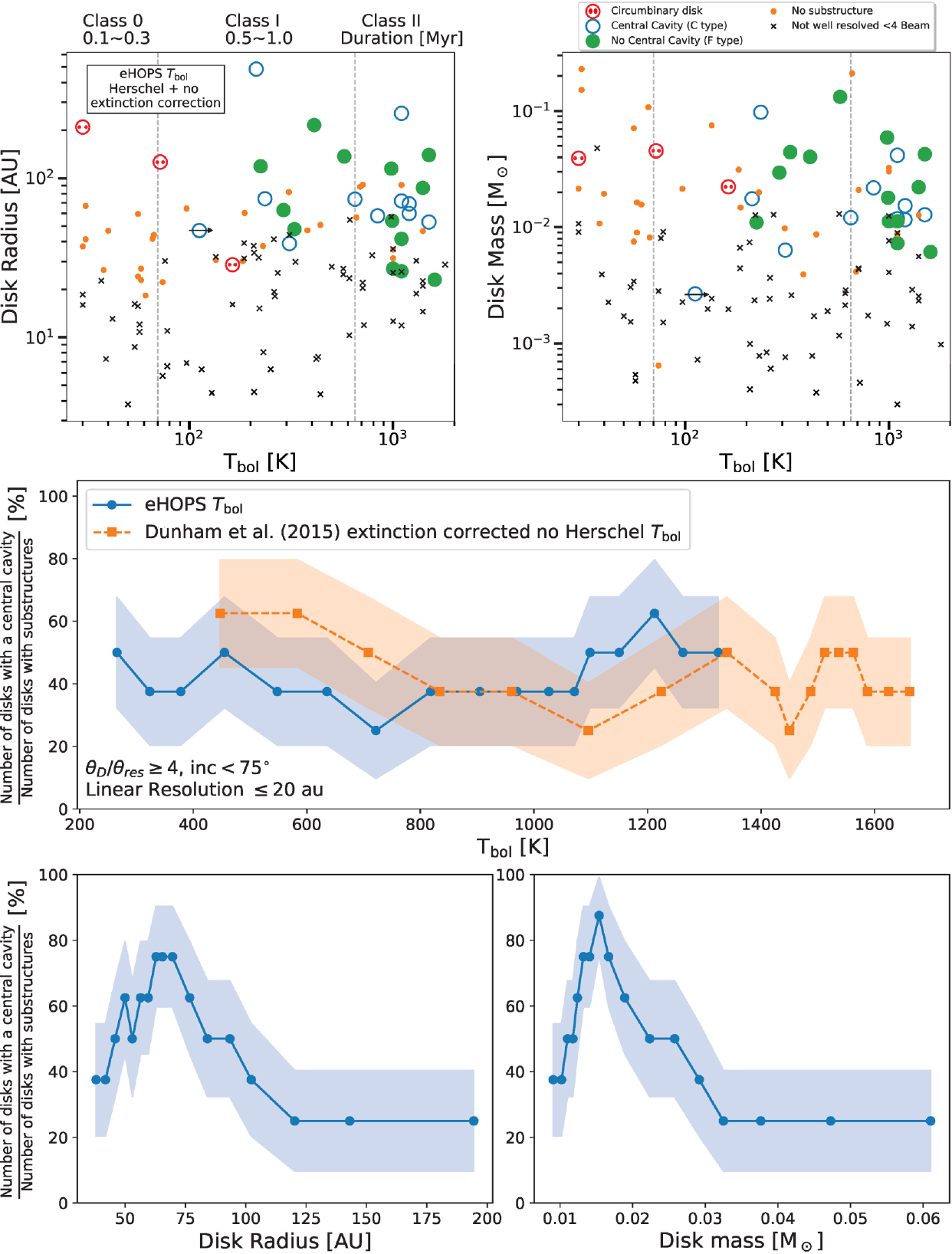}
    \caption{Upper panel: Protostellar disk substructure with or without a central cavity as a function of disk radius, mass, and bolometric temperature. Middle panel: Fraction of disk with a central cavity versus the evolutionary stage traced by bolometric temperature. Bottom panel: Fraction of disk with a central cavity versus the disk mass and radius. The bin size of 8 disks is used to compute the fraction. The shaded region represents the $1\sigma$ uncertainty from counting statistics. All 24 protostellar disks are observed by ALMA Band 6 at 1.3\,mm wavelength, have a disk diameter greater than 4 times the resolution ($\theta_D/\theta_{res} \ge 4$), with an inclination angle less than 75\,$^\circ$, have substructures, and linear resolution less than 20\,au.} 
\label{fig:CvsF_type}
\end{figure*}

\subsection{The protostellar disk substructures with and without central cavities}

We found that some protostellar disks have a central cavity in Class I and flat-spectrum phase. To further investigate this, we divide the substructure into two more general groups. The cavity-type (C-type) and the centrally filled type (F-type). The cavity-type disk substructures exhibit large ($>$10 au) cleared central dust cavities. These C-type protostellar disk substructures display morphologies similar to the more evolved Class II/Class III transition disks \citep{2016A&A...592A.126V}. The origin of the central dust cavity in transition disks can be attributed to a range of processes, like extensive particle growth \citep{2005ApJ...625..414T,2005A&A...434..971D}, photoevaporative mass-loss in winds \citep{2001MNRAS.328..485C}, dead zones \citep{2012MNRAS.419.1701R} or embedded stellar or planetary companions \citep{2008ApJ...678L..59I}. In particular, in systems with super-Jovian planets or stellar binaries that can clear the material in disks, deep gas gaps are expected inside cavities in the disk dust distribution \citep{2018MNRAS.477.1270P,2019ApJ...879L...2M,2021AJ....161...33V}. The other type of disk substructure population is the centrally filled type (F-type) disk substructure. In contrast to the C-type disk substructures, F-type disk substructures do not exhibit a central cavity.

Figure \ref{fig:2_population} shows that these 2 types of substructures as a function of evolutionary state. C-type disk substructures are found across all evolutionary stages from Class I to Class II. The detection of large cavities in younger Class I and Flat-spectrum sources cannot be explained by photoevaporative mass-loss in winds \citep{2001MNRAS.328..485C}. Instead, the central cavity in Class I can possibly be explained by radial drift \citep{2010A&A...513A..79B,2017MNRAS.465.3865C}. In this scenario, dust particles experience a headwind due to the gas drag, drifting towards the central protostar \citep{2017MNRAS.465.3865C}. Another explanation for these central cavities is tidal interactions with companions. 
The identification of binary systems in ALMA images relies on the presence of circumstellar disks around both components. While our ALMA data does not show evidence of binarity within the cavity of the C-type disks shown in Figure~\ref{fig:2_population}, we cannot exclude the presence of tight binary companions that either lack detectable dust disks or remain unresolved due to the linear resolution limit of 15\,au. Protostellar disks with C-type disk substructures are good candidates to search for gap-opening protoplanets or close protobinaries.

Notably, we found that multiple-ring systems, which have at least two rings with similar brightness within a factor of two at different disk radii, are almost always centrally filled (F-type). Among the 13 multiple-ring systems complied from the literature  \citep{2018ApJ...869L..41A,2021MNRAS.501.2934C,2020Natur.586..228S,2024ApJ...973..138H,2015ApJ...808L...3A,2023Natur.623..705S,2018ApJ...869L..42H}, we found that the majority (10 out of 13 or $77\,\%$) are centrally filled (F-type), as shown in Figure \ref{fig:Multiple_rings}. The three exceptions (HD 142666, HD 143006, and Sz 129) are among the oldest systems \citep{2018ApJ...869L..42H}. Notably, the central cavities in these systems are significantly smaller than their disk sizes \citep{2001MNRAS.328..485C}. Multiple-ring systems with a large central cavity ($\ge 25$\,\% of the disk diameter) are absent or very rare. The high fraction of F-type multiple-ring systems implies that a centrally filled disk is a general feature of multiple-ring systems. 

In Figure \ref{fig:CvsF_type}, we examine the prevalence of central cavities (C-type) in protostellar disks as a function of disk mass, disk radius, and evolutionary stage, as indicated by the bolometric temperature. For disks with substructures, each disk is assigned a value of 1 if it exhibits a central cavity (C-type) and 0 if it does not (F-type). We compute a moving average with a window size of 8, where the kernel slides through $T_{\rm bol}$, disk radius, and disk mass, and calculate the mean for every group of 8 disks. This mean represents the fraction of disks with a central cavity (C-type). 

Figure \ref{fig:CvsF_type} shows that, in general, protostellar disks with a central cavity are smaller and less massive than those without a central cavity. Classifying disks with substructures into those with and without large central cavities, we find both populations coexisting across evolutionary stages. The fraction of protostellar disks with a central cavity stays roughly constant at around $40\%$.

\section{Discussion}
\label{sec:discussion}

\subsection{Central cavities in protostellar disks: are disk substructures evolutionarily linked?}
\label{sec:dis:CvsFtype}

\citet{2021MNRAS.501.2934C} proposed that the diversity of disk substructures is evolutionarily linked and can be explained by giant planet formation via core accretion, coupled with simple dust evolution. Building upon this framework, \citet{2025ApJ...984L..57O} conducted numerical simulations of protoplanetary disks to further establish the unified evolutionary sequence of planet-induced substructures. The evolutionary scenario proposed by both studies outlines five distinct stages, defined primarily by morphological features.

Classifying disk morphologies serves as a critical first step in examining this evolutionary sequence, analogous to the Hubble tuning fork used for galaxy classification. However, unlike galaxy classification, which relies on easily distinguishable features such as spiral versus elliptical structures, differentiating between stages of annular substructures is considerably more challenging due to their subtler morphological distinctions.

To test whether or not protostellar disk substructures are evolutionarily linked, as argued by \citet{2021MNRAS.501.2934C} and \citet{2025ApJ...984L..57O}, we adopt a binary classification scheme based on the presence (C-type) or absence (F-type) of a central cavity. This classification is advantageous due to its simplicity, robustness, and clarity. If disk substructures indeed evolve over time, as proposed, we expect F-type disks to dominate in the earlier stages, with an increasing prevalence of C-type disks at later evolutionary stages.

By classifying disks with substructures into those with (C-type) and without (F-type) central cavities, we find that both populations coexist across evolutionary stages (Figure~\ref{fig:CvsF_type}). Our results suggest that there is no clear, uniform evolutionary sequence in disk substructures. If the evolution of the disk substructures follows the unified evolutionary sequence proposed by \citet{2021MNRAS.501.2934C} and \citet{2025ApJ...984L..57O}, then the data suggest that disk substructures evolve very rapidly and thus can be present in all Class I/II stages and/or that they can be triggered at different times. Alternatively, rather than representing an evolutionary path, the observed diversity could be explained by multiple distinct populations of disk substructures, shaped by different formation mechanisms.

We found that disks with a central cavity are, in general, smaller in size and less massive than disks without a central cavity. Most multi-ring systems fall into the F-type category, and multi-ring systems with a large central cavity (occupying at least 25\% of the disk diameter) are exceedingly rare. A few notable exceptions include ISO-Oph 2a \citep{2024ApJ...973..138H,2020ApJ...902L..33G} and HD 169142 \citep{2019AJ....158...15P}, though these are atypical cases. HD 169142, for instance, is significantly older ($\sim$10 Myr; \citealt{2017ApJ...850...52P}) than the protostellar disks analyzed in this study (ages $\lesssim 1$ Myr). This disk also hosts an embedded Jupiter-mass protoplanet, which interacts with the disk and likely contributes to the formation of a central cavity \citep{2023ApJ...952L..19L}. The other exception, ISO-Oph 2a, when observed at higher angular resolution, reveals two concentric rings \citep{2020ApJ...902L..33G}. Recent multi-frequency ALMA observations show that this disk exhibits azimuthal temperature variations likely caused by a fly-by interaction with a secondary companion \citep{2023MNRAS.526.1545C}. The morphology of ISO-Oph 2a, shaped by dynamical interactions, differs significantly from that of the other multi-ring systems presented in \autoref{fig:Multiple_rings}.

We speculate that the two populations of disk substructures are related to different planetary systems architectures, formed through different mechanisms, and will possibly evolve into different types of planetary systems. The recent discovery of the first protoplanetary system, PDS\,70, revealed two giant planets located at 20 and 35 au from the host star \citep{2018A&A...617A..44K,2019NatAs...3..749H}. In contrast, our solar system is much more compact, with Jupiter at $\sim5$\,au and Saturn at $\sim10$\,au. Jupiter formed much closer to the Sun at $\sim 3$\,au before entering a resonance with Saturn and eventually migrating to its current position, as suggested in the Grand Track Model \citep{2011Natur.475..206W}. The formation of Jupiter-like giant planets would create a strong pressure bump around the planet, blocking mm-sized dust grains and preventing the dust in the outer disk from replenishing the inner disk. Over time, this causes the inner disk to dissipate completely, forming a central cavity \citep{2012ApJ...755....6Z,2016A&A...585A..35P}. Thus, protostellar disks with C-type substructures might form compact planetary systems that resemble our Solar system. 

On the other hand, the gaps and rings in the multiple-rings system imaged by ALMA (F-type) are located much further away, at 10s of au from the host star. The location of these gaps and rings contrasts with the distribution of giant planets' orbits, which peaks at 3\,au as measured by the California Legacy Survey \citep{2021ApJS..255...14F}. These multiple-ring systems are also larger and more massive than the disks with a central cavity, possibly suggesting that they might be the precursors of extended planetary systems like PDS\,70. Alternatively, simulations show that multiple gaps and rings can be generated by a single planet inside a low viscosity disk \citep{2017ApJ...843..127D,2018ApJ...869L..47Z} or explained by various instabilities (see \citealt{2023ASPC..534..423B} for a review). We hypothesize that some of the F-type multiple-ring disks with shallower gaps originated from various instabilities or low-mass planets inside a low-viscosity disk. In contrast, disks with a central cavity (C-type) are more likely to be associated with giant planet formation as compared to disks without a central cavity.

\subsection{The onset of protostellar disk substructures}
\label{sec:dis:onset_substructures}

\begin{figure*}[tbh!]
    \includegraphics[width=1\textwidth]{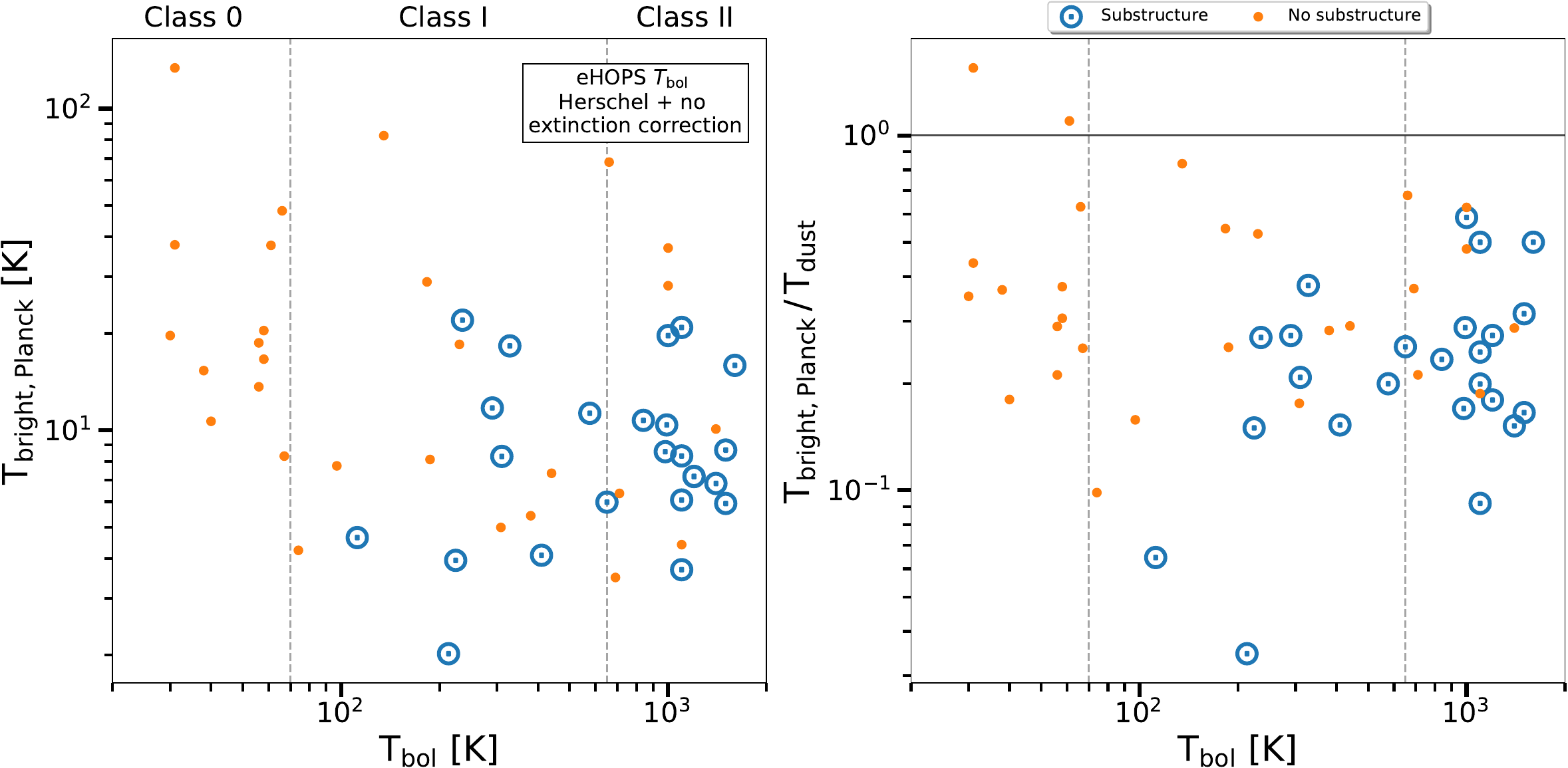}
    \caption{Left: The brightness temperature of the protostellar disks calculated from the full Planck equation as a function of the evolutionary stage (bolometric temperature). 
    Right: The transition between disks with substructures and without substructures at $T_{\rm bol}\sim200$\,K is not due to the optical depth. We do not see a discontinuity for the brightness temperature to dust temperature ratio ($T_{\rm bright,Planck} \, / \, T_{\rm dust}$) at 200\,K. For optically thick disks the $T_{\rm bright,Planck} \, / \, T_{\rm dust}$ ratio should be close to 1 (black horizontal line). We adopted the dust temperature from Equation 1, and excluded the circumbinary disks.} 
\label{fig:Optical_depth}
\end{figure*}

Since the discovery of rings and gaps in the HL Tau protoplanetary disk \citep{2015ApJ...808L...3A}, many more evolved protoplanetary disks substructures (rings, gaps, spirals) have been detected (e.g., \citealt{2018ApJ...869...17L,2019ApJ...871....5T,2017ApJ...839...99P,2017A&A...607A..55V,2016A&A...585A..58V,2018ApJ...868L...5K,2018ApJ...859...32P,2018ApJ...869L..48G,2018A&A...616A..88V,2017A&A...600A..72F,2022ApJ...941..172L,2024AJ....167..164B,2024ApJ...966...59S,2024ApJ...976..132H,2025A&A...696A.232G}). \citet{2025A&A...696A.232G} recently observed over 73 Class II disks in the Lupus molecular cloud at a resolution of 4\,au, and found that the substructure detection rate strongly depends on the effective resolution. For disks with sizes between 8 and 30\,au ($2 \le \theta_{\rm D}/\theta_{\rm res}\le 7.5$), the substructure detection rate is approximately 40\,\%, whereas it increases to $\sim$80\% for disks larger than 30\,au ($\theta_{\rm D}/\theta_{\rm res}\ge 7.5$) \citep{2025A&A...696A.232G}. Although many Class II disks are compact \citep{2025A&A...696A.232G}, the majority of Class II protoplanetary disks observed at high linear (a few au) and effective resolution ($\theta_{\rm D}/\theta_{\rm res}\ge 7.5$) exhibit substructures \citep{2018ApJ...869L..41A,2018ApJ...869L..42H,2023ASPC..534..423B,2025A&A...696A.232G}. The high substructure detection rate in well-resolved Class II disks indicates that substructures are already widespread by the Class II stage \citep{2023ASPC..534..423B}.

The quest to search for disk substructures in the Class 0/I protostellar phase has generated a great interest in the disk community \citep{2021ApJ...920...71A,2023ApJ...954..190X,2023ApJ...951L...2L,2024A&A...689L...5M,2023AJ....166..184M,2023ApJ...951....8O}. In particular, the recent ALMA Large Program, eDisk, has made a significant contribution, surveying around 20 embedded Class 0/I protostellar disks and found that three of the disks in their sample have substructures (L1489 IRS, Oph IRS 63 and IRAS 04169+2702) \citep{2023ApJ...951....8O}. Among these, substructures in Oph IRS 63 and L1489 IRS had been previously reported \citep{2020Natur.586..228S,2022ApJ...933...23O}. Based on the prevalence of Class 0/I disks without substructures (detection rate 3 out of 20), \citet{2023ApJ...951....8O} concluded that substructures form rapidly near the Class I/II transition.

Our analysis of 90 protostellar disks from CAMPOS, incorporating observations from eDISK, ODISEA, and DSHARP, shows that detection rates for Class I and Class II disks become comparable at $T_{\rm bol} \sim 200-400$\,K. In this study, we accounted for potential biases, including variations in observational resolution and source inclination. Lower resolution and higher inclination can artificially reduce the visibility of disk substructures, disproportionately affecting the detection of younger disks in current high-resolution surveys. We also consider the possible systematic biases, such as the lack of extinction correction or Herschel data, on the bolometric temperature of each source. Our finding is consistent with the recent study of protostellar disks in the Ophiuchus molecular cloud by \citet{2025PASJ..tmp...33S}, which found that disk substructure starts to appear when $T_{\rm bol}$ exceeds 200–300\,K. The onset of disk substructure at $T_{\rm bol} \sim 200-400$\,K, corresponds to an approximate age of $\sim$0.2-0.4 Myr, which is much earlier than the age of $\sim 1\,$Myr -- the expected age for when the transition between the Class I and Class II takes place \citep{2009ApJS..181..321E,2015ApJS..220...11D}.

It is important to note that the onset of disk substructures in this study could be an upper limit, and the actual onset of disk substructures may occur much earlier, within the Class 0 phase. We took a conservative approach and did not count sources that show small power variations as a function of distance in $uv-$space or small variations in the intensity radial profile but appear smooth in the image plane as disks with substructures (e.g., TMC-1A spiral\footnote{We did not include TMC-1A as a substructure source because it was not identified as a disk with a spiral substructure in the eDisk survey \citep{2023ApJ...951....8O}, in contrast to \citet{2023ApJ...954..190X}.}: \citealt{2021ApJ...920...71A,2023ApJ...954..190X}, HH211 spiral\footnote{HH211 is not included in Table 1, because it is an edge-on source.}: \citet{2023ApJ...951L...2L}, Oph A SM1 gap\footnote{Oph A SM1 is not included in  Table 1 because it does not have a bolometric temperature measurement.}: \citet{2024A&A...689L...5M}, faint rings in Oph-emb-6, Oph-emb-9, and GSS 30 IRS 3\footnote{Ring-like structures are not visually identifiable by the naked eye in Oph-emb-6, Oph-emb-9, and GSS 30 IRS 3 \citep{2023AJ....166..184M,2024ApJ...973..138H}.}: \citet{2023AJ....166..184M}). 
Our conservative approach minimizes the chances of false positives, but we may miss lower-contrast (shallower) substructures in younger disks with borderline detection. Furthermore, another possible reason for this onset being an upper limit is the choice to include all disks with $\theta_D / \theta_{\rm res} \ge 4$, since the detection rate is only $\sim50\%$ for values between $3 < \theta_D / \theta_{\rm res} \le 10$ \citep{2023ASPC..534..423B}.

In addition, optical depth can also affect the detection rate of substructures in younger Class 0 or early Class I disks \citep[e.g.,][]{2024arXiv240720074M}. To investigate this possibility further, we estimated the optical depth at 1.3 mm of our disks by computing the ratio between brightness temperature ($T_{\rm B, Planck}$) and the dust temperature. We estimated the brightness temperature of the disk from the average disk intensity using the full Planck equation. The estimated brightness temperature as a function of evolution stage ($T_{\rm bol}$) is plotted in the left panel of Figure \ref{fig:Optical_depth}. We then compared it with the dust temperature given by:
\begin{equation}
T_{\mathrm{dust}}=43\left(\frac{L_{\mathrm{bol}}}{1 L_{\odot}}\right)^{0.25}
\end{equation}
, where $L_{\mathrm{bol}}$ is the bolometric luminosity
\citep{2020ApJ...890..130T}. The optical depth ($\tau_\nu$) follows the following relationship:
\begin{equation}
T_{\mathrm{bright, Planck}} = T_{\mathrm{dust}} \left(1-e^{-\tau_\nu}\right)
\end{equation}
For optically thick disks, the $T_{\rm bright,Planck} \, / \, T_{\rm dust}$ ratio is expected to be close to 1. In the right panel or Figure \ref{fig:Optical_depth}  we plot the $T_{\rm bright,Planck} \, / \, T_{\rm dust}$ ratio as a function of evolutionary stage.

Although we do not find a sharp transition in optical depth between Class 0 and I, there is an indication of optical depth increasing from Class I back into the Class 0 stage. The median brightness temperature to dust temperature ratio ($T_{\rm bright,Planck} / T_{\rm dust}$) for Class 0 sources is 0.36, which is higher than that of Class I sources ($70 \le T_{\rm bol} \le 650$\,K), at 0.23, and Class II sources ($T_{\rm bol} \ge 650$\,K), at 0.27. This supports that the zero detection rate in our large sample for $T_{\rm bol} \le 163$\,K might be affected by high optical depths at 1.3 mm. Further suggestive of this is the fact that the disks with detected substructures in our work seem to have, on average, less estimated optical depth than all the remaining disks with no substructures during the Class 0 and I stages. If the Class 0 protostellar disks are optically thick, then the onset of protostellar disk substructures likely occurs even earlier.

The rapid formation of disk substructures at early Class I phases suggests that protostars and protoplanets grow and evolve together. During the Class 0 and early Class I phases, the central protostar and protostellar disks are constantly fed by anisotropic infall via streamers \citep{2020NatAs...4.1158P,2022A&A...667A..12V,2023ASPC..534..233P}. This anisotropic infall can perturb the protostellar disks, generating vortices and pressure bumps, resulting in substructure formation \citep{2022ApJ...928...92K}. These earliest disk substructures may be triggered by anisotropic infall \citep{2022ApJ...928...92K,2025arXiv250117857Z}. Other mechanisms for substructure formation have also been proposed (see \citet{2023ASPC..534..423B} for a review). While we do not know the origin of these early disk substructures, these structures can possibly trigger subsequent giant planet formation \citep{2024A&A...688A..22L}. The early onset of protostellar disk substructures at $\sim$0.2 -- 0.4 Myr is also consistent with recent population synthesis models, which show that rapid substructure formation within 0.4 Myr is needed to explain the observed spectral index distribution in Lupus protoplanetary disks \citep{2024A&A...688A..81D}. Our findings of the early formation of disk substructures at early Class I phases confirm that star formation sets the initial conditions for planet formation in protostellar disks.

\section{Conclusion}
\label{sec:conclusion}

In summary, our ALMA study reveals that by $T_{\rm bol} \sim 200-400\,K$ protostellar disk substructure detection rates increase sharply to $\sim$60\,\%. Our statistical analysis confirms that substructures similar to those in Class II disks are already common by the Class I stage. We argue that this emergence is likely only an upper age limit, as our study suggests increasing optical depth toward the earliest stages.

The prevalence of substructures in the early protostellar stage suggests that planetesimals and giant planets form much more efficiently and earlier than predicted by traditional models (e.g., \citet{1996Icar..124...62P}). This early planet formation scenario is also supported by a low abundance of water in the inner Solar System \citep{2016Icar..267..368M,2019A&A...629A..90H}, the existence of several isotopic reservoirs in the Solar system \citep{2017PNAS..114.6712K} as well as the distribution of calcium-aluminum-rich inclusions \citep{2018ApJS..238...11D,2020NatAs...4..492B}. Additionally, older protoplanetary disks (around Class II sources) might not have enough dust mass to form planetary systems \citep{2018A&A...618L...3M}. Our study has pushed the onset of planet formation to the much earlier protostellar phase, in $\sim$0.2--0.4 Myr disks. During this time period, the central protostar still has a high accretion rate and is building up its mass. 

In addition, by classifying disks with substructures into those with (C-type) and without (F-type) central cavities, we find that both populations coexist across evolutionary stages. If protostellar disk substructures are evolutionarily linked, then the data suggest that disk substructures evolve very rapidly and thus can be present at both Class I and Class II stages, and/or that they can be triggered at different times. Alternatively, the observed diversity could be explained by multiple distinct populations of disk substructures, shaped by different formation mechanisms.

\section*{Data Availability}
All data used in this study are public and can be downloaded from the ALMA Science Archive. \autoref{table:CAMPOS_Tbol}, \autoref{table:substructure_data}, and \autoref{table:ODISEA2019_Tbol} are also available in electronic form at the CDS via anonymous ftp to cdsarc.u-strasbg.fr (130.79.128.5) or via http://cdsweb.u-strasbg.fr/cgi-bin/qcat?J/A+A/.

\begin{acknowledgements}
C.H.H. and H. A are supported in part by NSF grants AST-1714710. C.H.H. is supported by the NASA Hubble Fellowship Program under award HF2-51556. D.S.C. is supported by an NSF Astronomy and Astrophysics Postdoctoral Fellowship under award AST-2102405. Special thanks to the e-HOPS group, Dr. Riwaj Pokhrel, and Dr. Thomas Megeath for providing the unpublished e-HOPs catalog for protostar classification. C.H.H. thanks Dr. Sean Andrews for the suggestion of considering the effective resolution in constructing a homogeneous sample for this statistical study. We also want to thank the referee for the insightful comments and discussion. This paper makes use of the following ALMA data: ADS/JAO. ALMA \#2016.1.00484.L, \#2018.1.00028.S, \#2019.1.00261.L, \#2019.1.01792.S. ALMA is a partnership of the ESO (representing its member states),NSF(USA),and NINS (Japan), together with the NRC (Canada), NSC and ASIAA(Taiwan), and KASI(Republic of Korea), in cooperation with the Republic of Chile. The Joint ALMA Observatory is operated by the ESO, AUI/NRAO, and NAOJ. The National Radio Astronomy Observatory is a facility of the National Science Foundation operated under cooperative agreement by Associated Universities, Inc. This paper makes use of the following software: Astropy: \citet{2013ascl.soft04002G,2018AJ....156..123A,2022ApJ...935..167A}, CASA: \citet{2007ASPC..376..127M}, SciPy:  \citet{2020NatMe..17..261V}.

\end{acknowledgements}

\bibliographystyle{aa}
\bibliography{Disk.bib}

\appendix

\section{Supplementary materials}
\label{Appendix}

In Table \ref{table:substructure_data}, we present all the disks in this study compiled from the CAMPOS survey \citep{2024ApJ...973..138H}, eDisk survey \citep{2023ApJ...951....8O}, DSHARP survey \citep{2018ApJ...869L..41A}, and ODISEA survey \citep{2021MNRAS.501.2934C,2019MNRAS.482..698C}. The protostellar systems in this study have bolometric temperatures less than 1900\,K, and inclination angles less than $75^\circ$. The Table consists of a homogeneous survey of 5 molecular clouds, Corona Australis, Chamaeleon I \& II, Ophiuchus \& Ophiuchus North, with additional sources in Taurus, Lupus I and BHR\,71 from the literature. 

In Table \ref{table:CAMPOS_Tbol}, we cross-match all the Corona Australis, Chamaeleon I \& II, Ophiuchus North, and Ophiuchus sources in our CAMPOS survey with the Young Stellar Objects catalog from the Spitzer Space Telescope ``cores to disks" (c2d) and ``Gould Belt" (GB) Legacy surveys \citep{2015ApJS..220...11D}, as well as the Extension of HOPS Out to 500 ParSecs (eHOPS) catalog to obtain the bolometric temperature ($T_{\rm bol}$), which serves as a proxy for relative evolutionary age for embedded protostellar systems. The eHOPS catalog contains 1-850 $\mu$m SEDs assembled from 2MASS, Spitzer, Herschel, WISE, and JCMT/SCUBA-2 data. It represents the latest and most reliable SED fitting to date.  The first eHOPS paper, covering the Serpens and Aquila molecular clouds, was published by \citet{2023ApJS..266...32P}. For all other clouds, the SED and the protostellar system properties are available in \href{https://irsa.ipac.caltech.edu/data/Herschel/eHOPS/overview.html}{NASA/IPAC Infrared Science Archive}.

In Table \ref{table:ODISEA2019_Tbol}, we cross-matched all the sources from the ODISEA survey \citep{2019MNRAS.482..698C} with the Spitzer Space Telescope ``cores to disks" (c2d) and ``Gould Belt" (GB) Legacy surveys \citep{2015ApJS..220...11D}, as well as the eHOPS survey \citep{2023ApJS..266...32P}, to obtain bolometric temperature and luminosity. Nine protostellar disks have bolometric temperature measurements but are unresolved in the ODISEA survey \citep{2019MNRAS.482..698C} and are therefore not included in Table \ref{table:substructure_data}.

\onecolumn

\begin{landscape}
\begin{longtable}{lrrrrrrrrrrrrrrr}
\caption{Protostellar disk substructures table} \\ %
\label{table:substructure_data} \\
\hline
Name & Class & eHOPS & Dun+15 & ${\rm L}_{\rm bol}$ &  Mass &  $R_{90\%}$ &  dist &  $F_{1.3mm}$ & Type & Beam & Res & $T_{\rm br,P}$ & $T_{\rm dust}$ & Ref \\
&& ${\rm T}_{\rm bol}$ [K] & ${\rm T}_{\rm bol}$ [K] & [$L_\odot$] & [$10^{-5} M_\odot$] & [au] & [pc] & [mJy] & &[mas]& [au] & [K] & [K] & \\
\hline
\endfirsthead

\multicolumn{15}{c}%
{{\tablename\ \thetable{} -- Continued from previous page}} \\
\hline
Name & Class & eHOPS & Dun+15 & ${\rm L}_{\rm bol}$ &  Mass &  $R_{90\%}$ &  dist &  $F_{1.3mm}$ & Type & Beam & Res & $T_{\rm br,P}$ & $T_{\rm dust}$ & Ref\\
&& ${\rm T}_{\rm bol}$ [K] & ${\rm T}_{\rm bol}$ [K] & [$L_\odot$] & [$10^{-5} M_\odot$] & [au] & [pc] & [mJy] & &[mas]& [au] & [K] & [K] & \\
\hline
\endhead

\hline
\multicolumn{15}{r}{\textit{Continued on next page}} \\
\endfoot

\hline
\endlastfoot

\multicolumn{15}{p{0.5\linewidth}} {Ophiuchus \& Ophiuchus North}\\
\hline
VLA 1623B & 0 & 30 & 30 & 2.87 & 21.48 & 37.3 & 144 & 114.2 & NA & 130 x 100 & 16.4 & 19.7 & 56.0 & 1 \\
VLA 1623Aa & 0 & 30 & 30 & 2.87 & 10.7 & 18.5 & 144 & 57.1 & NA & 130 x 100 & 16.4 & 35.2 & 56.0 & 1 \\
VLA 1623Ab & 0 & 30 & 30 & 2.87 & 9.09 & 15.9 & 144 & 48.3 & NA & 130 x 100 & 16.4 & 39.5 & 56.0 & 1 \\
VLA1623A & 0 & 30 & 30 & 0.45 & 39.22 & 209.7 & 144 & 208.5 & C & 130 x 100 & 16.4 & 4.1 & 35.2 & 1, 2, 3 \\
circumbinary & \\
IRAS 16293-2422A & 0 & 31 & 45 & 16.25 & 152.08 & 66.9 & 144 & 808.6 & NA & 62 x 50 & 8.0 & 37.7 & 86.3 & 1, 4 \\
IRAS 16293-2422B & 0 & 31 & 45 & 16.25 & 229.02 & 41.3 & 144 & 1217.8 & NA & 62 x 50 & 8.0 & 133.9 & 86.3 & 1, 4 \\
CB 68 SMM 1 & 0 & 38$^a$ & 39 & 0.89 & 10.75 & 26.5 & 151 & 52.0 & NA & 36 x 27 & 4.7 & 15.3 & 41.8 & 1, 5 \\
Oph-emb 1 & 0 & 42 & 36 & 0.19 & 2.26 & 13.1 & 144 & 12.0 & NA & 73 x 48 & 8.2 & 17.5 & 28.4 & 1, 5 \\
eHOPS-oph-20b & 0$^{b}$ & 57 & 830 & 1.77 & 0.54 & 12.1 & 144 & 2.9 & NA & 130 x 100 & 16.4 & 7.8 & 49.6 & 1 \\
eHOPS-oph-20c & 0$^{b}$ & 57 & 830 & 1.77 & 0.48 & 10.8 & 144 & 2.5 & NA & 130 x 100 & 16.4 & 8.2 & 49.6 & 1 \\
Oph-emb 6 & 0 & 67 & 120 & 0.35 & 8.18 & 44.0 & 144 & 43.5 & NA & 130 x 100 & 16.4 & 8.3 & 33.1 & 1 \\
Oph-emb 16 &  I  & 74 & 420 & 30.02 & 2.82 & 5.7 & 144 & 15.0 & NA & 130 x 100 & 16.4 & 87.8 & 100.7 & 1 \\
Oph-emb 9 &  I  & 76 & 200 & 0.33 & 8.05 & 30.2 & 144 & 42.8 & NA & 130 x 100 & 16.4 & 13.1 & 32.6 & 1 \\
Oph-emb 10a$^c$ &  I  & 78 & 180 & 10.25 & 9.12 & 11.0 & 144 & 48.5 & NA & 130 x 100 & 16.4 & 77.9 & 76.9 & 1 \\
Oph-emb 10b$^c$ &  I  & 78 & 180 & 10.25 & 1.52 & 6.6 & 144 & 8.1 & NA & 130 x 100 & 16.4 & 38.6 & 76.9 & 1 \\
ISO-Oph 31 & I & 97$^a$ & 250 & 1.7 & 21.42 & 64.5 & 138 & 124.0 & NA & 68 x 45 & 7.6 & 7.8 & 49.1 & 5 \\
Oph-emb 8 &  I  & 97 & 250 & 21.16 & 2.27 & 6.9 & 144 & 12.1 & NA & 130 x 100 & 16.4 & 50.9 & 92.2 & 1 \\
CFHTWIR-Oph 79\textsuperscript{\textdagger} &  I  & 112 & 300 & 7.78 & 2.67 & 46.9 & 144 & 14.2 & C & 56 x 39 & 6.7 & 4.6 & 71.8 & 1, 6 \\
Oph-emb 12 &  I  & 115 & 230 & 0.67 & 0.72 & 6.3 & 144 & 3.9 & NA & 130 x 100 & 16.4 & 22.5 & 38.9 & 1 \\
Oph-emb 13 &  I  & 129 & 280 & 11.11 & 1.98 & 4.5 & 144 & 10.5 & NA & 130 x 100 & 16.4 & 99.9 & 78.5 & 1 \\
Oph-emb 4 &  I  & 135 & 160 & 0.1 & 2.43 & 32.0 & 144 & 12.9 & NA & 130 x 100 & 16.4 & 6.2 & 24.2 & 1 \\
Oph-emb 11 &  Flat  & 163 & 260 & 0.71 & 1.97 & 16.1 & 144 & 10.5 & NA & 130 x 100 & 16.4 & 11.9 & 39.5 & 1 \\
ODISEA C4 064 & Flat & 185 & 1100 & 0.026 & 6.67 & 39.2 & 144 & 35.5 & NA & 280 x 190 & 33.2 & 8.5 & 17.3 & 7 \\
IR Cha INa4 &  I  & 186 & 430 & 0.06 & 4.43 & 31.4 & 179 & 15.2 & NA & 120 x 90 & 18.6 & 8.6 & 21.3 & 1 \\
ODISEA C4 102 & I & 207 & 207 & 0.13 & 7.43 & 37.6 & 144 & 39.5 & NA & 280 x 190 & 33.2 & 9.4 & 25.8 & 7 \\
IRAS 16237-2428 &  Flat  & 221 & 380 & 1.88 & 12.7 & 15.2 & 144 & 67.6 & NA & 130 x 100 & 16.4 & 58.3 & 50.4 & 1 \\
ISO-Oph-54 &  I  & 224 & 470 & 0.14 & 11.0 & 119.0 & 141 & 61.0 & F & 58 x 38 & 6.6 & 3.9 & 26.3 & 1, 8 \\
IRAS 16442-0930 &  I  & 230$^a$ & 280 & 0.44 & 19.96 & 37.5 & 144 & 106.1 & NA & 130 x 70 & 13.7 & 18.5 & 35.0 & 1 \\
Oph-emb 15 &  I  & 231 & 330 & 0.17 & 0.78 & 8.1 & 144 & 4.2 & NA & 130 x 100 & 16.4 & 16.4 & 27.6 & 1 \\
Oph-emb 24 &  Flat  & 251 & 560 & 0.29 & 0.84 & 6.3 & 144 & 4.5 & NA & 130 x 100 & 16.4 & 25.3 & 31.6 & 1 \\
ODISEA C4 081 & II$^d$ & 260 & 260 & 0.07 & 3.67 & 19.3 & 144 & 19.5 & NA & 280 x 190 & 33.2 & 14.2 & 22.1 & 7 \\
Oph-emb 18 &  I  & 263 & 430 & 0.04 & 0.61 & 18.1 & 144 & 3.2 & NA & 130 x 100 & 16.4 & 5.5 & 19.2 & 1 \\
ODISEA C4 033 & Flat & 271 & 540 & 0.204 & 12.8 & 25.4 & 144 & 67.8 & NA & 280 x 190 & 33.2 & 24.0 & 28.9 & 7 \\
ISO-Oph-17 &  II  & 290$^a$ & 1300 & 1.0 & 29.57 & 63.0 & 141 & 164.0 & F & 25 x 23 & 3.4 & 11.7 & 43.0 & 1, 8 \\
Oph-emb 19 &  Flat  & 310$^a$ & 450 & 0.48 & 0.76 & 44.0 & 144 & 4.1 & NA & 130 x 100 & 16.4 & 3.2 & 35.8 & 1 \\
Oph-emb 20 &  Flat  & 310 & 420 & 0.74 & 6.36 & 38.9 & 144 & 33.8 & C & 130 x 100 & 16.4 & 8.3 & 39.9 & 1 \\
Oph-emb 17 & Flat & 328 & 390 & 1.61 & 44.25 & 47.8 & 132 & 280.0 & F & 34 x 25 & 3.8 & 18.3 & 48.4 & 1, 5, 9 \\
ODISEA C4 107 & Flat & 332 & 332 & 0.03 & 2.6 & 29.8 & 144 & 13.8 & NA & 280 x 190 & 33.2 & 6.8 & 17.9 & 7 \\
ISO-Oph 93 &  Flat  & 380$^a$ & 550 & 0.04 & 3.93 & 47.0 & 144 & 20.9 & NA & 130 x 100 & 16.4 & 5.4 & 19.2 & 1 \\
Elias 2-27 &  Flat  & 410 & 1500 & 0.15 & 40.27 & 216.0 & 116 & 330.0 & F & 49 x 47 & 5.6 & 4.1 & 26.8 & 1, 7, 10, 11 \\
Oph-emb 21$^{e}$ &  I  & 420 & 520 & 3.3 & 0.78 & 7.3 & 144 & 4.2 & NA & 130 x 100 & 16.4 & 19.0 & 58.0 & 1 \\
Oph-emb 25 &  Flat  & 429 & 690 & 0.4 & 1.72 & 7.5 & 144 & 9.2 & NA & 130 x 100 & 16.4 & 34.3 & 34.2 & 1 \\
CFHTWIR-Oph 43 &  Flat  & 440$^a$ & 570 & 0.12 & 0.38 & 4.4 & 144 & 2.0 & NA & 130 x 100 & 16.4 & 23.9 & 25.3 & 1 \\
Oph-emb 23 &  Flat  & 440$^a$ & 570 & 0.12 & 8.68 & 50.8 & 144 & 46.2 & NA & 130 x 100 & 16.4 & 7.4 & 25.3 & 1 \\
Oph-emb 26a &  Flat  & 500$^a$ & 620 & 0.97 & 1.9 & 27.8 & 144 & 10.1 & NA & 130 x 100 & 16.4 & 6.3 & 42.7 & 1 \\
Oph-emb 26b &  Flat  & 500$^a$ & 620 & 0.97 & 1.89 & 27.7 & 144 & 10.1 & NA & 130 x 100 & 16.4 & 6.3 & 42.7 & 1 \\
F-MM7 &  Flat  & 570$^a$ & 690 & 0.72 & 1.17 & 27.0 & 144 & 6.2 & NA & 130 x 100 & 16.4 & 5.2 & 39.6 & 1 \\
Oph-emb 27 &  Flat  & 570$^a$ & 690 & 0.72 & 13.0 & 24.9 & 144 & 69.0 & NA & 130 x 100 & 16.4 & 25.2 & 39.6 & 1 \\
Oph-emb 28 &  Flat  & 610$^a$ & 720 & 1.1 & 2.13 & 10.3 & 144 & 11.3 & NA & 130 x 100 & 16.4 & 24.3 & 44.0 & 1 \\
ODISEA C4 086 & Flat & 610 & 720 & 1.1 & 2.7 & 23.5 & 144 & 14.3 & NA & 280 x 190 & 33.2 & 9.1 & 44.0 & 7 \\
ODISEA C4 045 & Flat & 613 & 613 & 0.35 & 2.87 & 54.7 & 144 & 15.2 & NA & 280 x 190 & 33.2 & 4.2 & 33.1 & 7 \\
ODISEA C4 026 & Flat & 690 & 1000 & 0.0023 & 4.15 & 88.4 & 144 & 22.1 & NA & 280 x 190 & 33.2 & 3.5 & 9.4 & 7 \\
ISO-Oph 37 & Flat & 710 & 810 & 0.24 & 20.92 & 91.0 & 141 & 116.0 & NA & 27 x 24 & 3.6 & 6.4 & 30.1 & 8 \\
ODISEA C4 043 & II & 790 & 1600 & 0.092 & 1.74 & 32.8 & 144 & 9.2 & NA & 280 x 190 & 33.2 & 5.2 & 23.7 & 7 \\
EM*SR 24S &  II  & 840 & 1300 & 1.3 & 21.83 & 58.0 & 115 & 182.0 & C & 37 x 33 & 4.0 & 10.7 & 45.9 & 8 \\
ODISEA C4 144 & II & 980 & 1500 & 0.0011 & 1.48 & 57.1 & 144 & 7.8 & NA & 280 x 190 & 33.2 & 3.3 & 7.8 & 7 \\
Elias 2-24 &  II  & 980$^a$ & 1500 & 1.9 & 59.05 & 115.0 & 136 & 352.0 & F & 37 x 34 & 4.8 & 8.6 & 50.5 & 1, 7, 10, 11 \\
Elias 2-20 &  II  & 990$^a$ & 1600 & 0.5 & 17.96 & 54.0 & 138 & 104.0 & F & 32 x 23 & 3.7 & 10.4 & 36.2 & 1, 7, 10, 11 \\
ODISEA C4 055 & II & 1000 & 1600 & 0.083 & 4.09 & 12.6 & 144 & 21.8 & NA & 280 x 190 & 33.2 & 29.7 & 23.1 & 7 \\
ODISEA C4 070 & II & 1000 & 1600 & 0.22 & 7.66 & 36.0 & 144 & 40.7 & NA & 280 x 190 & 33.2 & 10.1 & 29.4 & 7 \\
ODISEA C4 104 & II & 1000 & 1500 & 0.37 & 12.5 & 25.5 & 144 & 66.6 & NA & 280 x 190 & 33.2 & 23.5 & 33.5 & 7 \\
WSB 52 & II & 1000 & 1500 & 0.37 & 11.24 & 27.0 & 136 & 67.0 & F & 33 x 27 & 4.1 & 19.7 & 33.5 & 7, 10, 11 \\
ISO-Oph 2a &  II  & 1100$^a$ & 1600 & 0.11 & 11.85 & 72.0 & 144 & 63.0 & C & 30 x 21 & 3.6 & 6.1 & 24.8 & 1, 8 \\
ISO-Oph 2b &  II  & 1100$^a$ & 1600 & 0.11 & 0.3 & 11.9 & 144 & 1.6 & NA & 130 x 100 & 16.4 & 5.9 & 24.8 & 1, 8 \\
DoAr 20 &  II  & 1100$^a$ & 1400 & 0.89 & 7.3 & 41.5 & 144 & 38.8 & F & 130 x 100 & 16.4 & 8.3 & 41.8 & 1 \\
EM* SR4 & II & 1100 & 1400 & 0.89 & 11.24 & 26.0 & 134 & 69.0 & F & 34 x 34 & 4.6 & 20.9 & 41.8 & 7, 10, 11 \\
WSB 82 & II & 1100 & 1500 & 0.76 & 41.62 & 256.0 & 155 & 191.0 & C & 31 x 24 & 4.2 & 3.7 & 40.1 & 8 \\
ISO-Oph 196 & II & 1200 & 1600 & 0.14 & 15.32 & 69.0 & 137 & 90.0 & C & 45 x 22 & 4.3 & 7.2 & 26.3 & 8 \\
DoAr 44 & II & 1200 & 1500 & 0.75 & 11.6 & 60.0 & 146 & 60.0 & C & 34 x 25 & 4.3 & 7.2 & 40.0 & 8 \\
ODISEA C4 060 & II & 1300 & 1700 & 0.13 & 2.9 & 30.3 & 144 & 15.4 & NA & 280 x 190 & 33.2 & 7.1 & 25.8 & 7 \\
ODISEA C4 121 & II & 1300 & 1600 & 0.14 & 1.4 & 18.8 & 144 & 7.5 & NA & 280 x 190 & 33.2 & 8.0 & 26.3 & 7 \\
ODISEA C4 118 & II & 1400 & 1700 & 0.17 & 2.55 & 21.0 & 144 & 13.6 & NA & 280 x 190 & 33.2 & 10.0 & 27.6 & 7 \\
ODISEA C4 005 & II & 1400 & 1600 & 0.42 & 5.61 & 22.6 & 144 & 29.8 & NA & 280 x 190 & 33.2 & 15.4 & 34.6 & 7 \\
ODISEA C4 018 & II & 1400 & 1600 & 0.13 & 2.33 & 14.5 & 144 & 12.4 & NA & 280 x 190 & 33.2 & 15.5 & 25.8 & 7 \\
WaOph 6 &  II  & 1400$^a$ & 1700 & 1.2 & 22.09 & 87.0 & 123 & 161.0 & F & 58 x 54 & 6.9 & 6.8 & 45.0 & 1, 10, 11 \\
RX J1633.9-2442 &  II  & 1500 & 1800 & 0.17 & 12.8 & 53.0 & 141 & 71.0 & C & 21 x 20 & 2.9 & 8.7 & 27.6 & 1, 8 \\
DoAr 25 &  II  & 1500$^a$ & 1700 & 0.48 & 42.49 & 140.0 & 138 & 246.0 & F & 41 x 22 & 4.1 & 5.9 & 35.8 & 1, 7, 10, 11 \\
DoAr 33 & II & 1600 & 1800 & 0.3 & 6.13 & 23.0 & 139 & 35.0 & F & 37 x 24 & 4.1 & 15.9 & 31.8 & 10, 11 \\
ODISEA C4 017 & II & 1800 & 1900 & 0.085 & 0.98 & 28.6 & 144 & 5.2 & NA & 280 x 190 & 33.2 & 4.6 & 23.2 & 7 \\

\hline
\multicolumn{14}{p{0.3\linewidth}} {Chamaeleon I \& II}\\
\hline
TIC 454291385 & I & 74 & 74 & 1.0 & 0.65 & 22.2 & 189 & 2.0 & NA & 54 x 35 & 8.2 & 4.2 & 43.0 & 1, 5 \\
IRAS 12500-7658 &  I  & 187 & 160 & 0.31 & 14.76 & 60.4 & 181 & 49.7 & NA & 149 x 81 & 19.9 & 8.1 & 32.1 & 1 \\
V* GM Cha &  I  & 219 & 300 & 2.47 & 2.35 & 31.7 & 179 & 8.1 & NA & 120 x 90 & 18.6 & 6.1 & 53.9 & 1 \\
IRAS 12553-7651 &  I  & 260 & 260 & 1.74 & 2.42 & 41.4 & 144 & 12.9 & NA & 130 x 100 & 16.4 & 6.0 & 49.4 & 1 \\
IRAS 11030-7702 &  I  & 308 & 510 & 0.19 & 9.77 & 81.8 & 179 & 33.6 & NA & 74 x 43 & 10.1 & 5.0 & 28.4 & 1 \\
ISO-ChaI 101 &  Flat  & 650$^a$ & 820 & 0.09 & 12.06 & 73.8 & 179 & 41.5 & C & 99 x 65 & 14.4 & 6.0 & 23.6 & 1 \\
V* DK Cha &  II  & 660$^a$ & 1500 & 30.0 & 210.59 & 56.7 & 181 & 708.7 & NA & 146 x 82 & 19.8 & 68.2 & 100.6 & 1 \\
V* HO Cha a &  Flat  & 710$^a$ & 900 & 0.97 & 4.35 & 20.4 & 179 & 15.0 & NA & 120 x 90 & 18.6 & 14.8 & 42.7 & 1 \\
V* HO Cha b &  Flat  & 710$^a$ & 900 & 0.97 & 4.43 & 22.0 & 179 & 15.3 & NA & 120 x 90 & 18.6 & 13.5 & 42.7 & 1 \\
ISO-ChaI 204 &  II  & 720$^a$ & 1100 & 0.18 & 0.46 & 11.9 & 179 & 1.6 & NA & 120 x 90 & 18.6 & 7.2 & 28.0 & 1 \\
ISO-ChaI 207 &  II  & 1100$^a$ & 1500 & 0.09 & 8.75 & 90.7 & 179 & 30.1 & NA & 100 x 65 & 14.4 & 4.4 & 23.6 & 1 \\
ISO-ChaI 237 &  II  & 1100$^a$ & 1600 & 0.37 & 8.9 & 25.8 & 179 & 30.6 & NA & 120 x 90 & 18.6 & 17.7 & 33.5 & 1 \\
Ass Cha T 1-15 &  II  & 1400$^a$ & 1700 & 0.45 & 12.7 & 46.5 & 179 & 43.7 & NA & 100 x 65 & 14.4 & 10.1 & 35.2 & 1 \\

\hline
\multicolumn{14}{p{0.3\linewidth}} {Corona Australis}\\
\hline
V* VV CrA A &  0$^{b}$  & 37 & 610 & 1.55 & 47.8 & 22.7 & 149 & 237.3 & NA & 110 x 80 & 14.0 & 94.5 & 48.0 & 1 \\
IRS 5N & 0 & 40 & 23 & 3.59 & 19.4 & 46.7 & 147 & 99.0 & NA & 50 x 30 & 5.7 & 10.7 & 59.2 & 1, 5 \\
CrAus7-mm & 0 & 54 & 310 & 37.75 & 1.54 & 16.2 & 149 & 7.7 & NA & 110 x 80 & 14.0 & 10.1 & 106.6 & 1 \\
IRS 7A & 0 & 54 & 310 & 37.75 & 3.05 & 8.7 & 149 & 15.2 & NA & 110 x 80 & 14.0 & 44.0 & 106.6 & 1 \\
CXO 34 & 0 & 56 & 310 & 21.13 & 3.42 & 15.7 & 149 & 17.0 & NA & 110 x 80 & 14.0 & 18.2 & 92.2 & 1 \\
IRS 7B-a & 0 & 56 & 210 & 5.1 & 71.03 & 59.6 & 152 & 339.0 & NA & 54 x 42 & 7.2 & 18.7 & 64.6 & 1, 5 \\
IRS 7B-b & 0 & 56 & 210 & 5.1 & 7.54 & 24.1 & 152 & 36.0 & NA & 54 x 42 & 7.2 & 13.7 & 64.6 & 1, 5 \\
IRAS 32 A & 0 & 58 & 66 & 2.58 & 16.33 & 27.0 & 150 & 80.0 & NA & 35 x 23 & 4.3 & 20.4 & 54.5 & 1, 5 \\
IRAS 32 B & 0 & 58 & 66 & 2.58 & 8.98 & 22.9 & 150 & 44.0 & NA & 35 x 23 & 4.3 & 16.6 & 54.5 & 1, 5 \\
SMM 2 &  I  & 72 & 15 & 1.09 & 45.38 & 126.5 & 149 & 225.4 & C & 110 x 80 & 14.0 & 6.8 & 43.9 & 1 \\
V* V710 CrA$^f$ & I & 135 & 270 & 28.08 & 75.27 & 30.7 & 149 & 373.8 & NA & 120 x 80 & 14.6 & 82.5 & 99.0 & 1 \\
IRS 5a &  Flat  & 208 & 130 & 2.71 & 0.99 & 4.5 & 149 & 4.9 & NA & 110 x 80 & 14.0 & 51.4 & 55.2 & 1 \\
IRS 5b &  Flat  & 208 & 130 & 2.71 & 0.4 & 34.1 & 144 & 2.1 & NA & 130 x 100 & 16.4 & 3.1 & 55.2 & 1 \\
IRS 2 &  I  & 235 & 390 & 12.99 & 97.57 & 74.3 & 149 & 484.6 & C & 110 x 80 & 14.0 & 22.0 & 81.6 & 1 \\
V* S CrA A &  II  & 1000$^a$ & 1500 & 3.5 & 32.51 & 31.4 & 149 & 161.4 & NA & 110 x 80 & 14.0 & 36.9 & 58.8 & 1 \\
V* S CrA B &  II  & 1000$^a$ & 1500 & 3.5 & 30.27 & 35.6 & 149 & 150.3 & NA & 110 x 80 & 14.0 & 28.2 & 58.8 & 1 \\
\hline
\multicolumn{14}{p{0.3\linewidth}} {Taurus}\\
\hline
IRAS 04166+2706 & 0 & 61 & 61 & 0.4 & 15.67 & 18.3 & 156 & 71.0 & NA & 49 x 37 & 6.6 & 37.6 & 34.2 & 5 \\
IRAS 04169+2702 & I & 163 & 163 & 1.5 & 22.29 & 28.6 & 156 & 101.0 & C & 48 x 37 & 6.6 & 23.8 & 47.6 & 5 \\
TMC-1A & I & 183 & 183 & 2.3 & 31.32 & 30.1 & 137 & 184.0 & NA & 32 x 21 & 3.6 & 28.9 & 53.0 & 5 \\
L1489 IRS & I & 213 & 213 & 3.4 & 17.59 & 485.2 & 146 & 91.0 & C & 105 x 78 & 13.2 & 2.0 & 58.4 & 5 \\
HL Tau & I & 576 & 576 & 3.0 & 132.27 & 137.2 & 140 & 744.1 & F & 35 x 22 & 3.9 & 11.3 & 56.6 & 12, 13 \\
\hline
\multicolumn{14}{p{0.3\linewidth}} {BHR 71}\\
\hline
BHR 71 IRS2 & 0 & 39 & 39 & 1.1 & 3.93 & 7.3 & 176 & 14.0 & NA & 70 x 50 & 10.0 & 56.0 & 44.0 & 5 \\
BHR 71 IRS1 & 0 & 66 & 66 & 10.0 & 107.88 & 41.7 & 176 & 384.0 & NA & 72 x 53 & 10.9 & 48.1 & 76.5 & 5 \\
\hline
\multicolumn{14}{p{0.3\linewidth}} {Lupus I}\\
\hline
IRAS 15398-3359 & 0 & 50 & 50 & 1.4 & 1.72 & 3.8 & 155 & 7.9 & NA & 43 x 36 & 6.1 & 87.1 & 46.8 & 5 \\

\hline
\hline
\end{longtable}
\tablefoot{\textsuperscript{\textdagger} CFHTWIR-Oph 79 is also known as GY 263 or SKS 3-48 \citep{2015ApJ...815....2W,2023ApJ...958...20N}. The classification of this source has been debated. \citet{2023ApJ...958...20N} and  \citet{2002ApJ...566..993A} classified this source as a Class II disk, while \citet{2015ApJ...815....2W} classified it as a Flat spectrum disk based on J, H, and K band. The source is close to a Class I binary system, Oph-emb 14 VLA 1\&2.  In the CAMPOS survey, we assume the same bolometric temperature and class for all sources in the close multiple systems.  

$^{a}$ In this paper, if the source has no corresponding eHOPS match, we adopt the non-extinction corrected bolometric temperature from \citet{2015ApJS..220...11D}.

$^b$ The SED and protostellar system properties are available in \href{https://irsa.ipac.caltech.edu/data/Herschel/eHOPS/overview.html}{NASA/IPAC Infrared Science Archive}. eHOPS classified the source as Class I, but given its bolometric temperature, it should be Class 0.  

$^c$ eHOPS photometry for this source is not centered at the coordinates of the source for the IRAS $1\,\mu$m image. This could result in the incorrect value of $T_{\rm bol}$. We assigned values from \citet{2015ApJS..220...11D}.

$^d$ The eHOPS group calculated the spectral index between Spitzer/IRAC 4.5\,$\mu$m and Spitzer/IRAC 24 $\mu$m, and follows \citet{1994ApJ...434..614G} for Class classification \citep{2023ApJS..266...32P}. ODISEA C4 081 or eHOPS-oph-22 shows a negative slope between these wavelengths, resembling a Class II by definition. However, the source also shows significant emission at longer wavelengths, resulting in a lower bolometric temperature. 

$^{e}$ eHOPS photometry for this source is not centered at the coordinates of the source for the IRAS $1-4\,\mu$m image. This could result in the incorrect value of $T_{\rm bol}$. We assigned values from \citet{2015ApJS..220...11D}.

$^{f}$ eHOPS group updated the SED fit for this source recently, the original values published in \citet{2024ApJ...973..138H} for this source were outdated. 

\textbf{eHOPS ${\rm T}_{\rm bol}$:} Bolometric temperature from eHOPS catalog. The bolometric temperature is derived from SED fitting with Herschel Space Telescope longer wavelength data, but the IR extinction correction is not applied.   \textbf{${\rm L}_{\rm bol}$:} Bolometric luminosity. \textbf{Mass:} Disk dust mass. \textbf{R$_{90\%}$:} Disk dust $R_{90\%}$ radius. \textbf{dist:} Source distance. \textbf{$F_{1.3mm}$:} 1.3\,mm flux. \textbf{Type:} Type of the disk substructure. NA represents that no disk substructure is detected. C represents a ring with a large central cavity. F indicates that the disk is centrally filled. \textbf{Dun+2015 ${\rm T}_{\rm bol}$:} Bolometric temperature from \citet{2015ApJS..220...11D} catalog. The bolometric temperature is derived from SED fitting without Herschel Space Telescope data, but the IR extinction correction is applied. \textbf{$T_{\rm br,P}$:} Brightness temperature of the disk calculated from the full Planck equation. \textbf{$T_{\rm dust}$:} Dust temperature of the disk adopting equation 1. \textbf{Ref:} References. 

\textbf{References:} (1) Sources observed in our CAMPOS survey. See also the CAMPOS data paper, \citet{2024ApJ...973..138H}, for source identification and radius measurements. (2) \citet{2013A&A...560A.103M} (3) \citet{2020ApJ...894...23H} (4) \citet{2022ApJ...941L..23M} (5) \citet{2023ApJ...951....8O} (6) \citet{2023ApJ...958...20N} (7) \citet{2019MNRAS.482..698C} (8) \citet{2021MNRAS.501.2934C} (9) \citet{2020Natur.586..228S} (10) \citet{2018ApJ...869L..41A}  (11) \citet{2018ApJ...869L..42H} (12) \citet{2015ApJ...808L...3A} (13) \citet{2018ApJ...869...59W}}

\end{landscape}

\begin{longtable}{lrrrrr}
\caption{Cross matching Corona Australis, Chamaeleon I \& II, Ophiuchus North, and Ophiuchus sources in our CAMPOS survey} \\ %
\label{table:CAMPOS_Tbol} \\
\hline
Name & CAMPOS ID &  SSTgbs/SSTc2d Names &  eHOPS Name &  ISO Name &  IRAS sources \\

\hline
\endfirsthead

\multicolumn{5}{c}%
{{\tablename\ \thetable{} -- Continued from previous page}} \\
\hline
Name & CAMPOS ID &  SSTgbs/SSTc2d Names &  eHOPS Name &  ISO Name &  IRAS sources \\
\hline
\endhead
\hline
\multicolumn{5}{r}{\textit{Continued on next page}} \\
\endfoot

\hline
\endlastfoot

\multicolumn{6}{p{0.5\linewidth}} {Ophiuchus \& Ophiuchus North}\\
\hline
IRAS 16442-0930 & OphN-01-0 & SSTgbs J1646582-093519 & - & - & IRAS 16442-0930 \\
IRAS 16459-1411 & OphN-02-0 & SSTgbs J1648456-141636 & - & - & IRAS 16459-1411 \\
/ WaOph 6 & \\
CB 68 SMM 1 & OphN-03-0 & SSTgbs J1657196-160923 & - & - & IRAS 16544-1604 \\
ISO-Oph 2a & Oph-01-0 & SSTc2d J162538.1-242236 & - & ISO-Oph   2 & - \\
ISO-Oph 2b & Oph-01-1 & SSTc2d J162538.1-242236 & - & ISO-Oph   2 & - \\
DoAr 20 & Oph-02-0 & SSTc2d J162556.1-242048 & - & ISO-Oph   6 & IRAS 16229-2413 \\
ISO-Oph 17 & Oph-03-0 & SSTc2d J162610.3-242055 & - & ISO-Oph  17 & - \\
Elias 2-20 & Oph-05-0 & SSTc2d J162618.9-242820 & - & ISO-Oph  24 & IRAS 16233-2421 \\
Oph-emb 8 & Oph-06-0 & SSTc2d J162621.3-242304 & eHOPS-oph-2 & ISO-Oph  29 & - \\
ISO-Oph 31 & Oph-06-1 & SSTc2d J162621.7-242250 & - & ISO-Oph  31 & - \\
/ GSS 30 IRS3 & \\
DoAr 25 & Oph-07-0 & SSTc2d J162623.7-244314 & - & ISO-Oph  38 & IRAS 16234-2436 \\
Elias 2-24 & Oph-08-0 & SSTc2d J162624.1-241613 & - & ISO-Oph  40 & IRAS 16233-2409 \\
Oph-emb 9 & Oph-09-0 & SSTc2d J162625.5-242302 & eHOPS-oph-4 & - & - \\
VLA 1623B & Oph-10-0 & SSTc2d J162626.4-242430 & - & - & - \\
VLA 1623Ab & Oph-10-1 & SSTc2d J162626.4-242430 & - & - & - \\
VLA 1623Aa & Oph-10-2 & SSTc2d J162626.4-242430 & eHOPS-oph-5 & - & - \\
VLA 1623W & Oph-10-3 & SSTc2d J162625.6-242429 & - & - & - \\
Oph-emb 22 & Oph-11-0 & SSTc2d J162640.5-242714 & eHOPS-oph-7/8 & ISO-Oph  54 & - \\
IRAS 16237-2428 & Oph-12-0 & SSTc2d J162644.2-243448 & eHOPS-oph-9 & ISO-Oph  65 & IRAS 16237-2428 \\
Elias 2-27 & Oph-13-0 & SSTc2d J162645.0-242308 & eHOPS-oph-10 & ISO-Oph  67 & - \\
Oph-emb 23 & Oph-14-0 & SSTc2d J162648.5-242839 & - & ISO-Oph  70 & - \\
CFHTWIR-Oph 43 & Oph-14-1 & SSTc2d J162648.4-242835 & - & - & - \\
DoAr 29 & Oph-15-0 & SSTc2d J162658.5-244537 & - & ISO-Oph  88 & - \\
Oph-emb 21 & Oph-16-0 & SSTc2d J162702.3-243727 & eHOPS-oph-12 & ISO-Oph  92 & - \\
ISO-Oph 93 & Oph-17-0 & SSTc2d J162703.0-242615 & - & ISO-Oph  93 & - \\
Oph-emb 6 & Oph-18-0 & SSTc2d J162705.2-243629 & eHOPS-oph-16 & ISO-Oph  99 & - \\
Oph-emb 20 & Oph-19-0 & SSTc2d J162706.8-243815 & eHOPS-oph-17 & ISO-Oph 103 & - \\
Oph-emb 16 & Oph-20-0 & SSTc2d J162709.4-243719 & eHOPS-oph-19 & ISO-Oph 108 & - \\
eHOPS-oph-20a & Oph-21-0 & - & eHOPS-oph-20 & ISO-Oph 121 & - \\
eHOPS-oph-20b & Oph-21-1 & - & eHOPS-oph-20 & - & - \\
eHOPS-oph-20c & Oph-21-2 & SSTc2d J162715.8-243843 & eHOPS-oph-20 & - & - \\
Oph-emb 11 & Oph-22-0 & SSTc2d J162717.6-242856 & eHOPS-oph-23 & ISO-Oph 124 & - \\
Oph-emb 28 & Oph-23-0 & SSTc2d J162721.5-244143 & - & ISO-Oph 132 & - \\
Oph-emb 12 & Oph-25-0 & SSTc2d J162724.6-244103 & eHOPS-oph-26 & ISO-Oph 137 & - \\
Oph-emb 14 VLA 1 & Oph-26-0 & SSTc2d J162726.9-244051 & eHOPS-oph-29 & ISO-Oph 141 & IRAS 16244-2434 \\
Oph-emb 14 VLA 2 & Oph-26-1 & SSTc2d J162726.9-244051 & - & ISO-Oph 141 & IRAS 16244-2434 \\
CFHTWIR-Oph 79 & Oph-26-2 & - & - & - & IRAS 16244-2432 \\
Oph-emb-13 & Oph-27-0 & SSTc2d J162728.0-243933 & eHOPS-oph-30 & ISO-Oph 143 & - \\
Oph-emb 19 & Oph-28-0 & SSTc2d J162728.4-242721 & - & ISO-Oph 144 & - \\
Oph-emb 26a & Oph-29-0 & SSTc2d J162730.2-242743 & - & ISO-Oph 147 & - \\
Oph-emb 26b & Oph-29-1 & SSTc2d J162730.2-242743 & - & ISO-Oph 147 & - \\
Oph-emb 24 & Oph-30-0 & SSTc2d J162737.2-244238 & eHOPS-oph-35 & ISO-Oph 161 & - \\
Oph-emb 27 & Oph-31-0 & SSTc2d J162739.8-244315 & - & ISO-Oph 167 & IRAS 16246-2436 \\
F-MM7 & Oph-31-1 & SSTc2d J162739.8-244315 & - & ISO-Oph 167 & IRAS 16246-2436 \\
Oph-emb 1 & Oph-32-0 & SSTc2d J162821.6-243623 & eHOPS-oph-42 & - & IRAS 16253-2429 \\
Oph-emb 18 & Oph-33-0 & SSTc2d J162857.9-244055 & eHOPS-oph-43 & - & - \\
Oph-emb 17 / IRS 63 & Oph-34-0 & SSTc2d J163135.6-240129 & eHOPS-oph-44 & - & IRAS 16285-2355 \\
Oph-emb 4 & Oph-35-0 & SSTc2d J163136.8-240420 & eHOPS-oph-46 & - & - \\
Oph-emb 25 & Oph-36-0 & SSTc2d J163143.8-245525 & eHOPS-oph-47 & ISO-Oph 200 & - \\
Oph-emb 15 & Oph-38-0 & SSTc2d J163152.5-245536 & eHOPS-oph-48 & ISO-Oph 203 & - \\
Oph-emb 10a & Oph-39-0 & SSTc2d J163201.0-245643 & eHOPS-oph-49 & ISO-Oph 209 & - \\
Oph-emb 10b & Oph-39-1 & SSTc2d J163201.0-245643 & - & ISO-Oph 209 & - \\
IRAS 16293-2422A & Oph-40-0 & - & eHOPS-oph-51 & - & IRAS 16293-2422 \\
IRAS 16293-2422B & Oph-40-1 & SSTc2d J163222.6-242832 & - & - & - \\
EDJ 1013 & Oph-41-0 & SSTc2d J163355.6-244205 & - & - & - \\
\hline
\multicolumn{6}{p{0.5\linewidth}} {Chamaeleon I \& II}\\
\hline
IRAS 11030-7702 & ChamI-01-0 & SSTgbs J1104227-771808 & eHOPS-cha-2 & ISO-ChaI  46 & IRAS 11030-7702 \\
TIC 454291385 & ChamI-02-0 & SSTgbs J1106464-772232 & eHOPS-cha-3 & - & -\\
/ Ced 110 IRS4B & &&&\\
ISO-ChaI 101 & ChamI-05-0 & SSTgbs J1107213-772211 & - & ISO-ChaI 101 & - \\
Ass Cha T 1-15 & ChamI-06-0 & SSTgbs J1107435-773941 & - & ISO-ChaI 112 & - \\
V* HO Cha a & ChamI-07-0 & - & - & ISO-ChaI 126 & - \\
V* HO Cha b & ChamI-07-1 & SSTgbs J1108029-773842 & - & ISO-ChaI 126 & - \\
V* GM Cha & ChamI-08-0 & SSTgbs J1109285-763328 & eHOPS-cha-7 & ISO-ChaI 192 & - \\
ChamI-9 mm & ChamI-09-0 & - & - & - & - \\
ISO-ChaI 204 & ChamI-09-1 & SSTgbs J1109461-763446 & - & ISO-ChaI 204 & - \\
ISO-ChaI 207 & ChamI-10-0 & SSTgbs J1109472-772629 & - & ISO-ChaI 207 & - \\
IR Cha INa4 & ChamI-11-0 & SSTgbs J1110033-763311 & eHOPS-cha-11 & - & - \\
ISO-ChaI 237 & ChamI-12-0 & SSTgbs J1110113-763529 & - & ISO-ChaI 237 & - \\
TIC 454329229 & ChamI-13-0 & SSTgbs J1111107-764157 & - & - & - \\
V* DK Cha & ChamII-01-0 & SSTc2d J125317.2-770710 & - & - & IRAS 12496-7650 \\
IRAS 12500-7658 & ChamII-02-0 & SSTc2d J125342.9-771511 & eHOPS-cha-12 & - & IRAS 12500-7658 \\
IRAS 12553-7651 & ChamII-03-0 & STc2d J125906.6-770740 & eHOPS-cha-13 & ISO-ChaII 28 & IRAS 12553-7651 \\

\hline
\multicolumn{6}{p{0.5\linewidth}} {Corona Australis}\\
\hline
V* S CrA B & CrAus-01-0 & - & - & ISO-CrA 116 & IRAS 18577-3701 \\
V* S CrA A & CrAus-01-1 & SSTgbs J1901086-365720 & - & ISO-CrA 116 & IRAS 18577-3701 \\
IRS 2 & CrAus-02-0 & SSTgbs J1901415-365831 & eHOPS-cra-2 & - & - \\
IRS 5a & CrAus-03-0 & SSTgbs J1901480-365722 & - & - & - \\
IRS 5b & CrAus-03-1 & SSTgbs J1901480-365722 & eHOPS-cra-3 & - & - \\
IRS 5N & CrAus-04-0 & SSTgbs J1901484-365714 & eHOPS-cra-4 & - & - \\
V* V710 CrA & CrAus-05-0 & SSTgbs J1901506-365809 & eHOPS-cra-6 & - & - \\
IRS 7A & CrAus-07-0 & SSTgbs J1901553-365721 & eHOPS-cra-7 & - & - \\
SMM1C & CrAus-07-1 & - & - & - & - \\
CrAus7-mm & CrAus-07-2 & - & - & - & - \\
CrAus8-mm1 & CrAus-08-0 & - & - & - & - \\
IRS 7B-a & CrAus-08-1 & SSTgbs J1901564-365728 & eHOPS-cra-8 & - & - \\
CXO 34 & CrAus-08-2 & J190155.76-365727.7 & - & - & - \\
IRS 7B-b & CrAus-08-3 & SSTgbs J1901564-365728 & - & - & - \\
SMM 2 & CrAus-09-0 & SSTgbs J1901585-365708 & eHOPS-cra-9 & - & - \\
IRAS 32 A & CrAus-10-0 & SSTgbs J1902586-370735 & eHOPS-cra-10 & ISO-CrA 182 & IRAS 18595-3712 \\
IRAS 32 B & CrAus-10-1 & - & - & ISO-CrA 182 & IRAS 18595-3712 \\
V* VV CrA A & CrAus-11-0 & SSTgbs J1903068-371249 & eHOPS-cra-11 & - & IRAS 18597-3717 \\
V* VV CrA B & CrAus-11-1 & - & - & - & IRAS 18597-3717 \\
\end{longtable}
\tablefoot{\textbf{CAMPOS ID:} Sources observed in our CAMPOS survey. See also the CAMPOS data paper \citep{2024ApJ...973..138H}. \textbf{SSTgbs/SSTc2d Names:} Spitzer Space Telescope Gould Belt Survey / Space Telescope core to disk Survey Names. See \citep{2015ApJS..220...11D} for the spectral energy distribution (SED) fitting and the bolometric temperature, and the bolometric luminosity of the source. \textbf{eHOPS Name:} Extension of HOPS Out to 500 ParSecs (eHOPS) catalog, represents the latest and most reliable SED fitting to date.  The eHOPS catalog contains 1-850 $\mu$m SEDs assembled from 2MASS, Spitzer, Herschel, WISE, and JCMT/SCUBA-2 data. The first paper of eHOPS for Serpens and Aquila molecular clouds was published by \citet{2023ApJS..266...32P}. For all other clouds, the SED and protostellar system properties are available in \href{https://irsa.ipac.caltech.edu/data/Herschel/eHOPS/overview.html}{NASA/IPAC Infrared Science Archive}. \textbf{ISO Name}: Infrared Space Observatory (ISO) source name. \textbf{IRAS Name}: Infrared Astronomical Satellite (IRAS) source name.}

\begin{longtable}{lrrrrr}
\caption{Cross matching ODISEA survey with eHOPS and SSTgbs/SSTc2d} \\ %
\label{table:ODISEA2019_Tbol} \\
\hline
ODISEA ID & SSTgbs/SSTc2d Names & eHOPS Name & Other Name &  Included? & Note \\

\hline
\endfirsthead

\multicolumn{6}{c}%
{{\tablename\ \thetable{} -- Continued from previous page}} \\
\hline
ODISEA ID & SSTgbs/SSTc2d Names & eHOPS Name & Other Name &  Included? & Note \\
\hline
\endhead
\hline
\multicolumn{5}{r}{\textit{Continued on next page}} \\
\endfoot

\hline
\endlastfoot
ODISEA C4 005 & J162218.5-232148 & - & - & Y & - \\
ODISEA C4 017 & J162506.9-235050 & - & - & Y & - \\
ODISEA C4 018 & J162524.3-242944 & - & - & Y & - \\
ODISEA C4 022 &  J162538.1-242236 & - & ISO-Oph-2 & Y & - \\
ODISEA C4 026 & J162546.6-242336 & - & - & Y & - \\
ODISEA C4 027 & J162556.1-242048 & - & EM* SR4 & Y & - \\
ODISEA C4 030 & J162610.3-242055 & - & ISO-Oph-17 & Y & - \\
ODISEA C4 033 & J162617.2-242345 & eHOPS-oph-1 & - & Y & - \\
ODISEA C4 034 & J162618.9-242820 & - & Elias 2-20 & Y & - \\
ODISEA C4 038 & J162623.5-242439 & - & ISO-Oph 37 & Y & - \\
ODISEA C4 039 & J162623.6-244314 & - & DoAr 25 & Y & - \\
ODISEA C4 041 & J162624.0-241613 & - & Elias 2-24 & Y & - \\
ODISEA C4 042 & J162625.4-242301 & eHOPS-oph-4 & Oph-emb 9 & Y & - \\
ODISEA C4 043 & J162627.5-244153 & - & - & Y & - \\
ODISEA C4 045 & - & eHOPS-oph-6 & - & Y & - \\
ODISEA C4 047 & J162640.4-242714 & eHOPS-oph-7 & ISO-Oph-54 & Y & - \\
ODISEA C4 051 & J162645.0-242308 & eHOPS-oph-10 & Elias 2-27 & Y & - \\
ODISEA C4 055 & J162648.9-243825 & - & - & Y & - \\
ODISEA C4 057 & - & eHOPS-oph-11 & - & N & unresolved \\
ODISEA C4 060 & J162656.7-241351 & - & - & Y & - \\
ODISEA C4 062 & J162658.5-244537 & - & EM*SR 24S & Y & - \\
ODISEA C4 064 & J162703.5-242005 & eHOPS-oph-13/14 & - & Y & - \\
ODISEA C4 065 & - & eHOPS-oph-15 & - & N & unresolved \\
ODISEA C4 067 & - & eHOPS-oph-16 & Oph-emb 6 & Y & - \\
ODISEA C4 068 & - & eHOPS-oph-17 & Oph-emb 20 & Y & - \\
ODISEA C4 070 & J162709.0-243408 & - & - & Y & - \\
ODISEA C4 071 & - & eHOPS-oph-18 & - & N & unresolved \\
ODISEA C4 080 & - & eHOPS-oph-21 & - & N & unresolved \\
ODISEA C4 081 & - & eHOPS-oph-22 & - & Y & - \\
ODISEA C4 082 & - & eHOPS-oph-23 & Oph-emb 11 & Y & - \\
ODISEA C4 086 & J162721.4-244143 & - & - & Y & - \\
ODISEA C4 087 & - & eHOPS-oph-24 & - & N & unresolved \\
ODISEA C4 090 & - & eHOPS-oph-26 & Oph-emb 12 & Y & - \\
ODISEA C4 091 & - & eHOPS-oph-28 & - & N & unresolved \\
ODISEA C4 099 & - & eHOPS-oph-35/36 & Oph-emb 24 & Y & - \\
ODISEA C4 102 & - & eHOPS-oph-37 & - & Y & - \\
ODISEA C4 104 & J162739.4-243915 & - & - & Y & - \\
ODISEA C4 105 & J162739.8-244315 & - & - & N & unresolved \\
ODISEA C4 107 & - & eHOPS-oph-38 & - & Y & - \\
ODISEA C4 111 & - & eHOPS-oph-40 & - & N & unresolved \\
ODISEA C4 112 & - & eHOPS-oph-41 & - & N & unresolved \\
ODISEA C4 114 & J162816.5-243658 & - & ISO-Oph 196 & Y & - \\
ODISEA C4 118 & J162854.0-244744 & - & - & Y & - \\
ODISEA C4 119 & - & eHOPS-oph-43 & Oph-emb 18 & Y & - \\
ODISEA C4 121 & J163023.3-245416 & - & - & Y & - \\
ODISEA C4 127 & J163133.4-242737 & - & DoAr 44 & Y & - \\
ODISEA C4 130 & J163135.6-240129 & eHOPS-oph-44 & Oph-emb 17 & Y & - \\
ODISEA C4 131 & - & eHOPS-oph-46 & Oph-emb 4 & Y & - \\
ODISEA C4 132 & J163143.7-245524 & eHOPS-oph-47 & Oph-emb 25 & Y & - \\
ODISEA C4 135 & - & eHOPS-oph-48 & Oph-emb 15 & Y & - \\
ODISEA C4 141 & J163355.6-244205 & - & RX J1633.9-2442 & Y & - \\
ODISEA C4 143 & J163945.4-240203 & - & WSB 82 & Y & - \\
ODISEA C4 144 & J163952.9-241931 & - & - & Y & - \\
\end{longtable}
\tablefoot{\textbf{SSTgbs/SSTc2d Names:} Spitzer Space Telescope Gould Belt Survey / Space Telescope core to disk Survey Names. See \citep{2015ApJS..220...11D} for the spectral energy distribution (SED) fitting and the bolometric temperature, and the bolometric luminosity of the source. \textbf{eHOPS Name:} Extension of HOPS Out to 500 ParSecs (eHOPS), is the latest, most reliable SED fitting to date. The eHOPS catalog contains 1-850 $\mu$m SEDs assembled from 2MASS, Spitzer, Herschel, WISE, and JCMT/SCUBA-2 data. The first paper of eHOPS for Serpens and Aquila molecular clouds was published by \citet{2023ApJS..266...32P}. For all other clouds, the SED and protostellar system properties are available in \href{https://irsa.ipac.caltech.edu/data/Herschel/eHOPS/overview.html}{NASA/IPAC Infrared Science Archive}. \textbf{Other Name}: Names shown in Table 1, other higher resolution data available.}

\twocolumn

\section{Bolometric temperature as a proxy for relative evolutionary age}
\label{Appendix_B} 
Tracing protostellar disk age is extremely difficult. For embedded protostars, the luminosity is dominated by accretion, resulting in a large spread in observed luminosity. This means individual protostellar ages cannot be directly determined from traditional isochrone fitting. Instead, the age sequence for protostellar disks is based on the dissipation of the protostellar envelope, as indicated by observables such as bolometric temperature ($T_{\rm bol}$) \citep{1995ApJ...445..377C,1993ApJ...413L..47M}. 

$T_{\rm bol}$ is the temperature of a blackbody having the same flux-weighted mean frequency as the observed continuum spectrum \citep{1995ApJ...445..377C}. As a protostar evolves toward the main sequence, the circumstellar envelope dissipates, shifting the $T_{\rm bol}$ to a higher temperature. We followed the standard method \citep{1989PASP..101..229W,2015ApJS..220...11D} to estimate the age for the onset of the protostellar disk substructures (at $T_{\rm bol} = 200$, $400$\,K).  The standard approach for determining the duration of each evolutionary class involves calculating the ratio of the number of sources in the evolutionary stage of interest to the number in a reference stage, and then multiplying this ratio by the duration of the reference evolutionary class \citep{2015ApJS..220...11D,1989ApJ...340..823W,2009ApJS..181..321E,2014prpl.conf..195D}. Following \citet{2009ApJS..181..321E}, \citet{2014prpl.conf..195D} and \citet{2015ApJS..220...11D}, we adopted Class II as the reference evolutionary class and assumed a Class II duration of 2 million years. We used the Spitzer Space Telescope cores to disks (c2d) and Gould Belt Legacy survey (gbs) catalog of 3000 young stellar objects in the 18 molecular clouds to estimate the age \citep{2015ApJS..220...11D}. We divided the number of protostars with bolometric temperatures less than $200$ and $400\,$K by the total number of protostars in Class 0, Class I, flat-spectrum phase, and Class II phase combined. We multiplied this ratio by 2 million years to obtain the corresponding age of 0.2--0.3 Myr, and 0.3--0.4 Myr, at $T_{\rm bol}= 200, 400$\,K, respectively.

Note that this commonly used number-counting method does not take into account different star-formation rates in different molecular clouds, assumes a steady state constant star formation rate, and treats the duration of each stage as a single value rather than a distribution \citep{2018A&A...618A.158K}. \citet{2009ApJS..181..321E} developed the idea that the protostellar lifetimes should be thought of as half-lives, in which they represent the timescales at which half of the objects will have moved through and left that $T_{\rm bol}$ range. \citet{2018A&A...618A.158K} accounted for a non-steady state star formation rate and adopted a sequential nuclear decay model to estimate the duration of each protostellar stage, and found that the half-life of Class 0 and Class I are $\sim4.7\times10^4$ and $\sim8.7\times10^4$ years, respectively.

If we adopt the duration from \citet{2018A&A...618A.158K}, the onset of disk substructure will occur much earlier. In this paper, we adopt a more conservative number-counting method for the age estimate. We also emphasize that the age estimate for the onset of protostellar disk substructures is only a zeroth-order estimate, and future work is needed to improve the $T_{\rm bol}$-age relation.

\end{document}